\documentclass[12pt]{article}
\usepackage{amsfonts}
\usepackage{amssymb}
\usepackage{latexsym}
\usepackage{graphicx}
\usepackage[english]{babel}
\begin{document}
\input epsf

\def\p{\partial}
\def\h{{1\over 2}}
\def\be{\begin{equation}}
\def\bea{\begin{eqnarray}}
\def\ee{\end{equation}}
\def\eea{\end{eqnarray}}
\def\d{\partial}
\def\la{\lambda}
\def\eps{\epsilon}
\def\bb{\bigskip}
\def\mm{\medskip}
\newcommand{\dm}{\begin{displaymath}}
\newcommand{\edm}{\end{displaymath}}
\renewcommand{\b}{\tilde{B}}
\newcommand{\gm}{\Gamma}
\newcommand{\ac}[2]{\ensuremath{\{ #1, #2 \}}}
\renewcommand{\ell}{l}
\newcommand{\z}{\ell}
\newcommand{\newsection}[1]{\section{#1} \setcounter{equation}{0}}
\def\bb{$\bullet$}

\def\q{\quad}

\def\bn{B_\circ}

\let\a=\alpha \let\b=\beta \let\g=\gamma \let\d=\delta \let\e=\epsilon
\let\c=\chi \let\th=\theta  \let\k=\kappa
\let\l=\lambda \let\m=\mu \let\n=\nu \let\x=\xi \let\r=\rho
\let\s=\sigma \let\t=\tau
\let\vp=\varphi \let\vep=\varepsilon
\let\w=\omega      \let\G=\Gamma \let\D=\Delta \let\Th=\Theta
                     \let\P=\Pi \let\S=\Sigma

\def\nn{\nonumber}
\let\bm=\bibitem

\let\pa=\partial

\begin{flushright}
\end{flushright}
\vspace{20mm}
\begin{center}
{\LARGE  Fuzzball geometries and higher derivative corrections for extremal holes}
\\
\vspace{18mm}
{\bf   Stefano Giusto\footnote{giusto@.mps.ohio-state.edu} and Samir D. Mathur\footnote{mathur@mps.ohio-state.edu}}\\

\vspace{8mm}
Department of Physics,\\ The Ohio State University,\\ Columbus,
OH 43210, USA\\
\vspace{4mm}
\end{center}
\vspace{10mm}
\thispagestyle{empty}
\begin{abstract}
2-charge D1-D5 microstates are described by geometries which end in `caps' near $r=0$; these caps reflect infalling quanta back in finite time. We estimate the travel time for 3-charge geometries in 4-D, and find agreement with the dual CFT. This agreement supports a picture of `caps' for 3-charge geometries. We argue that higher derivative corrections to such geometries arise from string winding modes. We then observe that the `capped' geometries have no noncontractible circles, so these corrections remain bounded everywhere and cannot  create a horizon or singularity. 

\end{abstract}
\newpage
\setcounter{page}{1}
\renewcommand{\theequation}{\arabic{section}.\arabic{equation}}
\section{Introduction}

The quantum physics of black holes presents some paradoxes. If black holes have no `hair' then how do we see the states that account for the Bekenstein entropy
$S_{Bek}={A\over 4G}$? Worse, if there is no  information about the state near the horizon then the Hawking radiation carries no data about the hole, and we lose unitarity \cite{hawking}.

Some  computations in string theory suggest that the conventional picture of the black hole interior is incorrect: Instead of `empty space with a central singularity'  we have a horizon sized `fuzzball', with state information distributed throughout the ball. The most explicitly studied system in this context is the 2-charge system described as follows. Take type IIB string theory with the compactification $M_{9,1}\rightarrow M_{4,1}\times T^4\times S^1$. Wrap $n_5$ D5 branes on $T^4\times S^1$ and $n_1$ D1 branes on $S^1$. The microscopic entropy of the bound state is $S_{micro}=2\sqrt{2}\pi\sqrt{n_1n_5}$. If one assumes that the D1-D5 bound state is pointlike in the noncompact $M_{4,1}$ then  one obtains the `naive metric' in which the $S^1$ shrinks to zero length as $r\rightarrow 0$, and horizon area vanishes. In \cite{lm4} the `actual' geometries of the system were constructed. At the classical level we get a continuous family of geometries; semiclassical quantization of this moduli space should yield the $e^{2\sqrt{2}\pi\sqrt{n_1n_5}}$ states in accordance with the microscopic count. These geometries depart from the `naive' geometry for $r$ smaller than some scale $r_s$.  Instead of  a singularity at $r=0$ the
throats of these geometries end in smooth `caps', with the shape of the cap differing from geometry to geometry. If we bound off this region $r<r_s$ by a surface, then the area of this bounding surface satisfies $A/4G\sim \sqrt{n_1n_5}\sim S_{micro}$
\cite{lm5}. 

If we add a momentum charge P along $S^1$ then we get the 3-charge D1-D5-P system where the naive geometry is a Reissner-Nordstrom type black hole. But for a subset of microstates of this system the actual geometries were constructed, and again the throat was found to end in caps \cite{mss,gms1,gms2}.\footnote{Similar geometries, but
with different values of momentum and angular momenta, were constructed in \cite{lunin}. It is not clear
to us if these geometries correspond to CFT microstates.} One would like to get more information about generic states of this system, since lessons learnt here are expected to extend to all black holes.

In this paper we show that some computations relating the CFT and gravity pictures for the 2-charge system also extend to the 3-charge system. We also argue that  higher derivative corrections for the 2-charge system (discussed recently in \cite{d, dkm, sen, hmr}) do not change the fact that the geometries end in `caps'. 

In more detail, we do the following:

\medskip

(a) \quad In 4+1 noncompact dimensions the naive geometry of the 2-charge D1-D5 extremal system has zero horizon area,
while the naive geometry of the 3-charge extremal D1-D5-P system has non-zero horizon area. Similarly in 3+1 noncompact dimensions the 3-charge extremal system has zero horizon area but the 4-charge extremal D1-D5-P-KK system has nonzero horizon area. 
To reach 3+1 noncompact dimensions we assume a compactification $M_{9,1}\rightarrow M_{3,1}\times T^4\times S^1\times \tilde S^1$. 

Suppose we take {\it three} charges D1-D5-P but compactify the additional direction $\tilde S^1$ as above. If the radius
$\tilde R$ of this $\tilde S^1$ is very large, then we have the usual 4+1 dimensional hole, whose microscopic entropy matches the Bekenstein entropy measured by the horizon area
$S_{micro}=2\pi\sqrt{n_1n_5n_p}=S_{Bek}$ \cite{stromvafa}. Now adiabatically reduce $\tilde R$ till it becomes much less than the horizon radius. At this stage we have  three charges in 3+1 dimensions. We can dualize  to make the three charges D1-D5-KK instead of D1-D5-P. Since all states are BPS we expect that their microscopic count will not change in this process. On the other hand we also expect that the area of the horizon should also not change in this process, so at the end we still expect  $S_{micro}=2\pi\sqrt{n_1n_2n_3}=S_{Bek}$ (we have relabeled the three charges  $n_1,n_2,n_3$). This suggests that the geometry for 3-charges in 3+1 dimensions should not be the naive one with zero horizon area. Instead we should have some kind of a `horizon' at $r=r_s$ where $r_s$ is such that if we place a surface at $r_s$ then its area will give ${A\over 4G_4}=2\pi\sqrt{n_1n_2n_3}$. This requirement yields 
\be
r_s\sim {G_4\over R}
\label{three}
\ee
(Here $R$ is the radius of $S^1$.)
\medskip

(b) \quad Next we recall the following fact about the D1-D5 system in 4+1 dimensions. The dual CFT is a 1+1 dimensional sigma model with target space $(T^4)^N/S_N$ (the symmetrized product of $N=n_1n_5$ copies of $T^4$). The orbifolding by $S_N$ generates twist sectors where $k$ copies of the CFT with target space  $T^4$ are linked up to give a similar CFT living on a `multiwound' circle that has length $(2\pi R)k$ instead of $2\pi R$. The CFT thus has many ground states, each of which is specified by the windings $k_i$ of its `component strings'.  

An incident graviton can be absorbed by the D1-D5 state, and in the CFT one left and one right moving excitation is created on a component string.  If the winding of the component string is $k$ then these excitations travel around the string in a time $\Delta t_{CFT}={(2\pi R)k\over 2}$ whereupon they can collide and reemerge from the string. In the dual {\it gravity} description, if we use the naive metric then we get a contradiction.
The `throat' down to $r=0$ is infinitely long, and quanta thrown down the throat fail to reemerge in the expected time. The contradiction is resolved when we realize that the actual geometries have `caps' which reflect the quanta back after a finite time. For states where component strings have the same length $(2\pi R)k$   we find an exact agreement of the travel times between the gravity and CFT pictures \cite{lm4}
\be
\Delta t_{sugra}={(2\pi R)k\over 2}=\Delta t_{CFT}
\ee
In the generic state contributing to the entropy not all the components have the same winding number, but we have
$\sim\sqrt{n_1n_5}$ components each with typical winding number $k_i\sim \sqrt{n_1n_5}$. The reflected wave from the cap is more complicated due to different phases arising from travel around components of somewhat different lengths, but we find  that the order of the time taken to reach the cap agrees with the order of travel time around the CFT component string \cite{lm4}
\be
\Delta t_{sugra}\sim \Delta t_{CFT}\sim \sqrt{n_1n_5}R
\label{one}
\ee

Now consider the {\it three} charge D1-D5-KK system in 3+1 dimensions discussed in (a) above. The CFT is very similar to the 2-charge case, with the total winding of the components equaling $n_1n_2n_3$. The generic state contributing to the entropy now has length $\sim\sqrt{n_1n_2n_3}R$, so the travel time for excitations will be 
\be
\Delta t_{CFT}\sim \sqrt{n_1n_2n_3}~R
\label{two}
\ee
This time we do not know all dual geometries, but if we had a picture similar to the 2-charge case then the `horizon' is located at the place where the generic geometry has its `cap'. But we have already argued in (a) above that the `horizon' should be at the location (\ref{three}). We compute the travel time $\Delta t_{sugra}$ in the geometry from the start of the throat down to the location $r_s$ and find that
\be
\Delta t_{sugra}\sim  \Delta t_{CFT}
\label{four}
\ee
This is one of the main observations of this paper. The agreement (\ref{four}) suggests that the entire picture of CFT states and dual `capped' geometries for the 2-charge D1-D5 system  also extends to 3-charge states.

\medskip

(c) \quad The capped geometries described above were found as solutions to the supergravity equations obtained at leading order in $\alpha'$ and $g$. It has recently been argued that for the D0-D4 system on $M_{3,1}\times K3\times T^2$ higher derivative terms correct the naive metric (which has zero horizon area) into a metric that has a finite horizon area.\footnote{Such  corrections were conjectured earlier in \cite{senold}, using suggestions in \cite{susskind}.}
Higher derivative corrections also modify the area law
for the entropy \cite{wald}, and in this case the correction is of
the same order as the area term. Thus the Bekenstein entropy gets modified from $A/4G$ to $A/2G$, which is found to
agree exactly with the microscopic count: ${A/ 2G}=S_{micro}$. 
How does this result relate to our picture where individual geometries have `caps' but no horizons?

The higher order terms used in \cite{d} were derived in the language of special geometry which describes the reduced action in 3+1 dimensions. We first identify the corresponding terms in  10-dimensional string theory; this enables us to locate similar effects in related cases like the D1-D5 system in 4+1 dimensions. In the string description these higher derivative terms arise from the following source. In the naive D1-D5 geometry the $S^1$ shrinks to zero size as $r\rightarrow 0$. Winding modes of the elementary string around this $S^1$ are thus very light for small $r$, and they give a  contribution to the string 1-loop effective action which diverges as $r\rightarrow 0$.

At first it might appear that this divergence would completely change the picture of `caps' that we had obtained without these correction terms, since if there is a divergent correction at some point then a smooth cap may well open up into a horizon. But in fact there is no location where such a divergent correction would occur. In the capped geometries the $S^1$ is
not actually a nontrivial circle \cite{lmm}. Just at the place where the correction from the winding modes would have become significant, this circle twists and mixes with the angular $S^3$ of the noncompact directions, and becomes the angular circle $\theta$ of plane polar coordinates $(r,\theta)$. Thus there is no winding mode at all, and we do not get any divergent contribution at $r=0$. A smooth cap is thus expected to remain a smooth cap, regardless of any metrical deformations that might arise from these higher derivative corrections.

\medskip

(d) \quad We  present some arguments to support the fuzzball picture of the black hole,  mention some puzzles about the role of higher derivative corrections, and close with a short discussion.  

\section{2-charge system: Review}

We summarize earlier results for the 2-charge D1-D5 system; these will be useful in the sections below.

Take IIB string with the compactification $M_{9,1}\rightarrow M_{4,1}\times T^4\times S^1$. Let the $S^1$ be parametrized by $y$ and the $T^4$ by $z_1\dots z_4$. The length of $S^1$ is $2\pi R$ and the volume of $T^4$ is $(2\pi)^4V$. The D1-D5 system is given by $n_5$ D5 branes wrapped on $T^4\times S^1$ and $n_1$ D1 branes wrapped on $S^1$. By S,T dualities we can map this to the FP system: The fundamental string $F$ is wrapped $n_5$ times around $S^1$ and carries $n_1$ units of momentum P along $S^1$. The {\it naive} metric of the FP bound state in string frame is ($u=t+y, v=t-y$)
\be
ds^2=(1+{Q_F\over r^2})^{-1}(-dudv+{Q_P\over r^2}dv^2)+dx_idx_i+dz_adz_a
\label{naivefp}
\ee
The dualities have changed the parameters $g, V, R$ to new values $g', V', R'$. In particular
\be
R=g'\sqrt{g\over V'}{\alpha'^2\over R'}
\label{vfive}
\ee
and we have
\be
Q_F={n_5 g'^2 \alpha'^3\over V'}, ~~~Q_P={n_1 g'^2 \alpha'^4\over V'R'^2}
\ee
But in the bound state of F and P  the momentum P is carried as traveling waves on the fundamental string F. The fundamental string has no longitudinal vibration modes so these vibrations are described by a transverse displacement profile $\vec{F}(v)$. We concentrate on the displacements in the noncompact directions $x^i$. The `actual' geometries of the FP system are determined by $\vec{F}(v)$ and in string frame are \cite{lm3,lm4}
\bea
ds^2&=&H(-dudv+K dv^2+2A_idx_i dv)+dx_idx_i+dz_adz_a\nonumber\\
B_{vu}&=&-G_{vu}=\h H, ~~B_{vi}=-G_{vi}=-H A_i, ~~e^{-2\Phi}=H^{-1}
\label{actualfp}
\eea
where
\be
H^{-1}=1+{Q_F\over L_T}\int_0^{L_T}\! {dv\over |\vec x-\vec F(v)|^2}, ~~K={Q_F\over
L_T}\int_0^{L_T}\! {dv (\dot
F(v))^2\over |\vec x-\vec F(v)|^2},
~~A_i=-{Q_F\over L_T}\int_0^{L_T}\! {dv\dot F_i(v)\over |\vec x-\vec F(v)|^2}\\
\label{functions}
\ee
Here 
\be
L_T=2\pi n_5 R',
\label{vtwo}
\ee
the total length of the F string.  Dualizing back to D1-D5 we get the string frame metrics
\be
ds^2=\sqrt{H\over 1+K}[-(dt-A_i dx^i)^2+(dy+B_i dx^i)^2]+\sqrt{1+K\over
H}dx_idx_i+\sqrt{H(1+K)}dz_adz_a
\label{six}
\ee
where after the dualities all lengths in the harmonic functions  get scaled by a factor 
\be
\mu={\sqrt{g\,V'}\over g'\,\alpha'}
\label{vthree}
\ee
 so that
\bea
H^{-1}=1+{\mu^2Q_F\over \mu L_T}\int_0^{\mu L_T} {dv\over |\vec x-\mu\vec F(v)|^2}, && ~~K={\mu^2Q_F\over
\mu L_T}\int_0^{\mu L_T} {dv (\mu^2\dot
 F(v))^2\over |\vec x-\mu\vec F(v)|^2},\nonumber\\
~~A_i=-{\mu^2Q_F\over \mu L_T}\int_0^{\mu L_T} {dv~\mu\dot F_i(v)\over |\vec x-\mu\vec F(v)|^2}&&
\label{functionsq}
\eea
Here $B_i$ is given by
\be
dB=-*_4dA
\label{vone}
\ee
and $*_4$ is the duality operation in the 4-d transverse  space
$x_1\dots
x_4$ using the flat metric $dx_idx_i$.
From the large distance behavior of $H^{-1}$ we see that
$\mu^2 Q_F$ must equal $Q_5$ where
\be
Q_5=n_5 g \alpha'
\label{vfour}
\ee  

\subsection{Regularity of the D1-D5 geometries}

At first sight it appears that the metrics (\ref{six}) have a singularity at points where $\vec x=\mu \vec F(v)$ for some $v$. But it was shown in \cite{lmm} that this is only a coordinate singularity; the geometries are completely smooth. To see this we compute the harmonic functions near such points. The points $\vec x=\mu \vec F(v)$ define a curve in the 4-dimensional Cartesian space spanned by the $x_i$; this space carries the flat metric $dx_idx_i$. Go to a point on the curve given by $v=v_0$. Let $z$ be a coordinate that measures distance along the curve (in the flat metric) and choose spherical polar coordinates $(\rho, \theta, \phi)$ for the 3-plane perpendicular to the curve. Then
\be
z\approx \mu |\dot {\vec F}(v_0)| (v-v_0)
\ee
\be
H^{-1}\approx {Q_5\over \mu L_T}\int_{-\infty}^\infty {dv\over \rho^2+z^2}={Q_5\over \mu L_T}{1\over \mu|\dot {\vec F}(v_0)|} \int_{-\infty}^\infty {dz\over \rho^2+z^2}={Q_5\pi\over \mu^2 L_T|\dot {\vec F}(v_0)|}{1\over \rho}
\ee
\be
K\approx {Q_5\pi |\dot {\vec F}(v_0)|\over  L_T}{1\over \rho}\,,\quad
A_z\approx -{Q_5\pi\over \mu L_T}{1\over \rho}
\ee
Let
\be
\tilde Q\equiv {Q_5\pi\over \mu L_T}
\ee
The field $A_i$ gives $F_{z\rho}=-{\tilde Q/ \rho^2}$, which gives from (\ref{vone}) $B_\phi=-
\tilde Q(1-\cos\theta)$. The $(y, \rho, \theta, \phi)$ part of the metric  becomes
\bea
ds^2&\rightarrow& \sqrt{H\over 1+K}(dy+B_i dx^i)^2+\sqrt{1+K\over
H}(d\rho^2+\rho^2(d\theta^2+\sin^2\theta d\phi^2))\nonumber\\
&\approx&{\rho\over \tilde Q}(dy-{\tilde Q}(1-\cos\theta) d\phi)^2+ {\tilde Q\over \rho}
(d\rho^2+\rho^2(d\theta^2+\sin^2\theta d\phi^2))
\label{vfourt}
\eea
This is the metric near the core of a Kaluza-Klein monopole, and is smooth if the length of the $y$ circle is $4\pi \tilde Q$, which implies
\be
 2\pi R =4\pi ({Q_5\pi\over \mu L_T})
\ee
Using (\ref{vfive}),(\ref{vtwo}),(\ref{vthree}),(\ref{vfour}) we see that this relation is exactly satisfied.

The $(t,z)$ part of the geometry gives
\bea
ds^2&\rightarrow&-\sqrt{H\over 1+K}(dt-A_zdz)^2+\sqrt{1+K\over
H}dz^2\nonumber\\
&\approx&-{\rho\over \tilde Q}dt^2-2 dtdz\approx -2dtdz
\eea
which is regular. The $T^4$ part gives $\mu|\dot {\vec F}(v_0)| dz_a dz_a$ and is thus regular as well. 

\subsection{Travel time}

\subsubsection{The D1-D5 CFT}

The low energy dynamics of a large number of D1,D5 branes is conjectured to be given by a 1+1 dimensional CFT which is  a sigma model with target space $(T^4)^{N}/S_{N}$ -- the symmetric product of $N=n_1n_5$ copies of $T^4$.
Each copy of $T^4$ gives a $c=6$ CFT with 4 free bosons, 4 free left moving fermions and 4 free right moving fermions. The orbifolding gives rise to twist sectors, created by twist operators $\sigma_k$ which link together $k$ copies of the $c=6$ CFT (living on  circles of length $2\pi R$) to give a $c=6$ CFT living on a circle of length $2\pi kR$. We call each such set of linked copies a `component string'.

The fermions in the CFT are in the Ramond  sector. On each component string we have 4 left and 4 right fermion zero modes, which give 8 bosonic and 8 fermionic ground states for the component string. The number of ways to partition $n_1n_5$ copies of the CFT into component strings, together with this zero mode degeneracy, reproduces the entropy $S_{micro}=2\sqrt{2}\pi\sqrt{n_1n_5}$. The fermion zero modes are in 1-1 correspondence with the forms $h_{p,q}$ on $T^4$. If we replace $T^4$ by K3 we will get a different number of zero modes, and the entropy $S=4\pi\sqrt{n_1n_5}$. 

By looking at the Poincare polynomials one can establish a direct link between the FP and D1-D5 states. A state of the FP string is generated by left oscillators acting on a ground state of the F string
\be
|\psi\rangle_{FP}=\alpha^{i_1}_{-k_1}\dots \alpha^{i_s}_{-k_s}|0\rangle
\label{veight}
\ee
In the D1-D5 CFT this maps to a Ramond ground state with component strings of lengths $k_1, \dots k_s$
\be
|\psi\rangle_{D1D5}=[\sigma_{k_1}^{a_1 b_1}\dots \sigma_{k_s}^{a_s b_s}|0\rangle]_{NS\rightarrow R}
\label{vnine}
\ee
where $|0\rangle$ is the NS vacuum and $NS\rightarrow R$ denotes that we spectral flow from the NS sector  to the R sector.
 The polarizations $i_1, \dots i_s$ of the oscillators map to the indices $\{a_i,b_i\}$ on the twist operators which denote the fermion zero modes applied to the component string.  The elements 
 \be
h_{00}, h_{20}, h_{02}, h_{22}
\label{vseven}
\ee
 of the cohomology in the D1-D5 system map to oscillators $\alpha^i_{-k}$ with polarization $i$ in the four noncompact directions $x_1\dots x_4$. Thus the solutions (\ref{six})  describe CFT states which have the zero modes (\ref{vseven}) on the component strings.\footnote{The oscillations of the F string in the $T^4$ directions were studied in \cite{lmm}.}
  
  Using  the map (\ref{veight}) $\rightarrow$ (\ref{vnine}) we can find the D1-D5 geometry that we want to associate (in the classical limit) with a Ramond ground state of the D1-D5 CFT. From the twist orders $k_m$ and polarizations $i_m$ we find the oscillator excitations  of the F string in the FP dual. These oscillators give, in the classical limit,  a profile function ${\vec F}(v)$. But for a given ${\vec F}(v)$  we know the FP geometry, and by duality the D1-D5 geometry
(\ref{six}).  

\subsubsection{Travel time for a special subclass}

Consider the following  special subset of states. Take $\vec F(v)$ of the form
\be
\vec F(v)=a[\sin(kv)\hat x_1+\cos(kv)\hat x_2]
\label{vttwo}
\ee
so that the F string makes a uniform helix of $k$ turns when opened up to the covering space. The corresponding CFT state turns out to be 
\be
|\Psi^{--}(k,0)\rangle=[(\sigma_k^{--})^{m}|0\rangle]_{NS\rightarrow R}
\label{vtthree}
\ee
where $m=N/k$ and  the superscripts $(--)$ on $\sigma_k$ indicate the fermion zero modes. Consider a graviton $h_{ab}$
incident on such a D1-D5 bound state ($a,b$ are indices on $T^4$). The graviton can be absorbed by the bound state, creating (at leading order) one left and one right moving excitation on one of the component strings (Fig.1(a), (b)). Thus the total state of the CFT has the form
\bea
|\psi\rangle_{excited}&=&{1\over \sqrt{m}}[{\rm component~string~1~excited}+{\rm component~string~2~excited}+\nonumber\\
&&~~~~~\dots+{\rm component~string~m~excited}]
\label{vtfour}
\eea
The length of each component string is $2\pi Rk$. The excitations travel around the string in opposite directions and collide at point B (Fig.2(c)) after a time
\be
\Delta t_{CFT}={2\pi Rk\over 2}=\pi Rk
\label{vtone}
\ee
Thus the graviton can reemerge from the bound state after a time $\Delta t_{CFT}$. If the collision does not result in emission then the excitations travel around the component string and again collide after a further time $\Delta t_{CFT}$ and so on.

\begin{figure}[ht]
\hspace{0in}
\includegraphics*[scale=.6, clip=false]{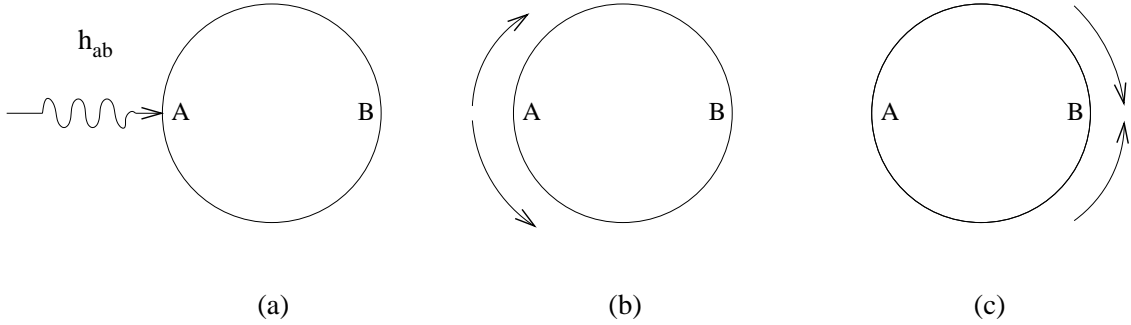}
\caption{\small{(a) A graviton is incident on the component string; (b) The graviton is absorbed at A and converted to a  pair of
vibration modes; (c) The vibration modes meet again at B.}}
\label{fig}
\end{figure}

We can construct the FP geometry for the profile (\ref{vttwo}) and thus the corresponding D1-D5 geometry:\footnote{These metrics were constructed earlier in \cite{balmm}.}
\bea
ds^2 & = & -\frac{1}{h} (dt^2-dy^2) + h f \left( \frac{dr^2}{r^2 +
a^2\,\gamma^2} + d\theta^2 \right)
\nonumber \\
         &+& h \Bigl( r^2 +
\frac{a^2\,\gamma^2\,Q_1 Q_5\,\cos^2\theta}{h^2 f^2} \Bigr)
\cos^2\theta d\psi^2  \nonumber \\
&+& h\Bigl( r^2 + a^2\,\gamma^2 -
\frac{a^2\,\gamma^2\,Q_1 Q_5 \,\sin^2\theta}{h^{2} f^{2} }
\Bigr) \sin^2\theta d\phi^2  \nonumber \\
&-& \frac{2a\,\gamma\,\sqrt{Q_{1}Q_{5}} }{hf}
(\cos^2\theta \,dy\,d\psi + \sin^2\theta \,dt\,d\phi)+\sqrt{H_1\over
H_5} \sum_{a=1}^4 dz_a dz_a
\label{mm}
\eea
where
\bea
&&a={\sqrt{Q_1 Q_5}\over R}\,,\quad
f=r^2+a^2\,\gamma^2\,\cos^2\theta\nonumber\\
&&H_1 =1+{Q_1\over f}\,,\quad H_5 =1+{Q_5\over f}\,,\quad  h =
\sqrt{H_1 H_5}
\eea
and $\gamma={1\over k}$ relates the choice of geometry to the corresponding CFT state (\ref{vtthree}). 

We can solve the wave equation for the graviton $h_{ab}$ in the geometry (\ref{mm}). We find that the graviton can be absorbed into the throat of the geometry (with exactly the same probability as the probability for absorption into the CFT bound state). The graviton then reflects off the `cap' at the end of the throat and a wavepacket exits the throat after a time delay 
\be
\Delta_{sugra}=\pi R k
\ee
We thus observe the exact agreement \cite{lm4}
\be
\Delta t_{CFT}=\Delta t_{sugra}
\label{vtseven}
\ee
 Part of the wave returns back down the throat and tries to reemerge after a further time $\Delta t_{sugra}$ etc, just as in the dual CFT.

\subsubsection{Travel times for generic states}

In a general CFT state (\ref{vnine}) the component strings have different lengths and different fermion zero modes. The shortest possible component string has winding number unity while the longest possible component string has winding number $n_1n_5$. In the generic state contributing to the entropy $S_{micro}$ the winding numbers are 
\be
k_i\sim \sqrt{n_1n_5}
\label{vttone}
\ee
which implies that there are $m\sim\sqrt{n_1n_5}$ component strings.\footnote{This fact can be seen from the FP picture where the winding numbers $k_i$ map to the harmonic orders $\alpha_{-k_i}$ of excitations. These vibrations form a 1-dimensional massless gas on the F string, which has total length $L_T=2\pi R n_5$.  The total energy of the vibrations on this string is $E={2\pi n_1n_5\over L_T}$. In the generic `thermal' state the temperature $T$ is given by $E\sim T^2L_T$, and the energy of the generic excitation is $\sim T$. These relations yield $k_i\sim \sqrt{n_1n_5}$.}

A graviton $h_{ab}$ incident on  a generic CFT state (\ref{vnine}) creates a state similar to (\ref{vtfour}). The excitations on the component strings separate, and then collide again after traversing the string as in Fig.1(b). But since the component strings have somewhat different lengths $2\pi R k_i$ the time of emission from different component strings is somewhat different. Thus while the average time of emission is
\be
\Delta t_{CFT}\sim \pi R\langle k\rangle 
\label{vtfive}
\ee
the emitted wavepacket is `distorted' compared to wavepacket of the incident graviton. A similar distortion is found in the dual gravity picture where the geometry now has a complicated `cap' and the graviton wave equation no longer separates between the radial and angular parts. Nevertheless we can still estimate the gravity travel time by computing how long the quantum takes to reach the region of the `cap'. It was shown in \cite{lm4} that the generic geometry can be approximated by the `naive' D1-D5 geometry 
\be
ds^2_{naive}={1\over \sqrt{(1+{Q_1\over r^2})(1+{Q_5\over
r^2})}}[-dt^2+dy^2]+\sqrt{(1+{Q_1\over r^2})(1+{Q_5\over
r^2})}dx_idx_i+\sqrt{{1+{Q_1\over r^2}\over 1+{Q_5\over r^2}}}dz_adz_a
\label{naive}
\ee
for $r\gtrsim r_s$, where
\be
r_s ={1\over\langle k\rangle}{\sqrt{Q_1Q_5}\over R}\sim {1\over \sqrt{n_1n_5}}{\sqrt{Q_1Q_5}\over R}
\label{vtsevenrs}
\ee
At the location $r\sim r_s$ the geometry turns into a complicated `cap'. The travel time for a massless radially infalling particle from the start of the throat to the location $r=r_s$ is
\be
\Delta t_{sugra}=\int _{r=r_s}^{(Q_1Q_5)^{1/4}}dr\,\sqrt{-{g_{rr}\over g_{tt}}} \approx 
{\sqrt{Q_1 Q_5}\over r_s}\sim R\sqrt{n_1 n_5}
\ee
We thus see that for generic states we get an analog of (\ref{vtseven})
\be
\Delta t_{CFT}\sim \Delta t_{sugra}
\label{vtsix}
\ee

\section{Travel times for 3-charge systems}

In this section we will get a relation similar to (\ref{vtsix}) for {\it three} charge states. 

Let us first outline the argument. For 2-charges D1-D5 in 4+1 dimensions we have the following facts. (i) The microstates
can be constructed from first principles, and are found to deviate from the naive geometry at some typical location $r\sim r_s$ and end in `caps'. (ii) The area of a boundary placed at $r=r_s$ in the naive metric gives $A/4G\sim \sqrt{n_1n_5}\sim S_{micro}$. (iii)  A quantum falls in from the start of the throat to $r\sim r_s$ in a time $\Delta t_{sugra}$ which agrees with the time $\Delta t_{CFT}$ for the excitations in the dual CFT to go around the component string; if we had instead the naive geometry extending down to $r=0$ we would get a contradiction because quanta would not return in the time expected from the CFT. 

Now we wish to consider 3-charge systems in 3+1 dimensions. We do not have the analogue of  (i) because we cannot yet construct the generic 3-charge microstates. But we argue that there should be an analogue of (ii), and since we know the entropy $\sim \sqrt{n_1n_2n_3}$ we can find the location $r=r_s$ for this case. This argument does not tell us whether at $r\sim r_s$ there should be  a horizon  or `caps'. Finally, we do the analogue of (iii), and find that the travel time to $r=r_s$ again matches the expectation from the CFT. This suggests that 3-charge geometries also end in caps.

\subsection{3-charge geometries in 3+1 dimensions}

In 4+1 noncompact dimensions we get a classical black hole with three charges, which we can take as D1-D5-P. In this case the microscopic entropy agrees exactly with the Bekenstein entropy \cite{stromvafa,callanmalda}
\be
S_{micro}=2\pi\sqrt{n_1 n_2 n_3}=S_{Bek}
\label{sv}
\ee
If we set the P charge to zero then we get the 2-charge system whose `naive' metric is given by (\ref{naive}) and has zero horizon area. But we have seen that the `actual' metrics (\ref{six}) depart from the naive metric and end in `caps'. The actual metrics start departing from the naive geometry at $r\sim r_s$ where $r_s$ is given in (\ref{vtsevenrs}). We can thus estimate the size of the `fuzzball' region if we place a boundary at
the location $r=r_s$ in the naive metric and compute the area of this boundary. This area $A$ turns out to satisfy a Bekenstein type relation \cite{lm5}
\be
{A\over 4G_5}\sim \sqrt{n_1n_2}\sim S_{micro}
\label{vteight}
\ee

If we compactify down to 3+1 noncompact dimensions then we can make  a classical hole with four charges, satisfying $S_{micro}=S_{Bek}$, just like (\ref{sv}). These charges can be taken to be D1-D5-KK-P, where KK stands for Kaluza-Klein monopoles \cite{jkm}. In the compactification $M_{3,1}\times T^4\times S^1\times \tilde S^1$ these KK monopoles extend uniformly along $T^4\times S^1$ while  $\tilde S^1$ is a nontrivially fibered circle. 

If we set the P charge to zero we get a `naive' metric  which has zero horizon area. In the 4-D Einstein frame this metric is
\bea
&&ds^2_{(E)}=-(H_1 H_2 H_3)^{-1/2}\,dt^2 + (H_1 H_2 H_3)^{1/2}\,(dr^2+d\Omega_2^2)\nonumber\\
&&H_i=1+{Q_i\over r}\,\,,\,\,i=1,2,3
\label{einstein3}
\eea
where we have relabeled the charges $Q_1, Q_2, Q_3$. 
This situation looks similar to that  with 2-charges in 4+1 dimensions, where the `actual' metrics departed from the naive ones at $r\sim r_s$. Should the 3-charge metrics have caps? 

\subsubsection{Locating the `horizon'}

Start with the D1-D5-P system in $M_{4,1}\times T^4\times S^1$ and compactify an additional direction to reach
$M_{3,1}\times T^4\times S^1\times \tilde S^1$.  When $\tilde S^1$ is large we  have (\ref{sv}). Now adiabatically reduce the size of $\tilde S^1$ till it is much smaller than the horizon radius of the D1-D5-P hole. At this point we have a 3-charge system in 3+1  dimensions. But since the states are all BPS, we still expect that the microscopic count gives $S_{micro}=2\pi\sqrt{n_1 n_2 n_3}$. What about the gravity description? There are some subtleties associated with the nature of the horizon after squeezing, which we discuss in subsection 3.3 below. But on general grounds we do not expect that the horizon area will shrink to zero as implied by the naive geometry (\ref{einstein3}). If the horizon continues to give the entropy, then we can find its location by putting a boundary at some $r=r_s$ in (\ref{einstein3}) and demanding that the
 area at this location give
\be
{A\over 4 G_4}={4\pi\sqrt{Q_1 Q_2 Q_3\,r_s}\over 4 G_4}=2\pi\sqrt{n_1 n_2 n_3}
\label{rss}
\ee
We do not know from this analysis whether the location $r=r_s$ is a conventional horizon or a boundary (in the sense of (\ref{vteight})) of the region where the typical caps occur. But we can compute the travel time from the start of the throat down to $r=r_s$:
\be
\Delta t_{sugra} = \int_{r_s}^{(Q_1 Q_2 Q_3)^{1/3}}\!dr\, \sqrt{-{g^{(E)}_{rr}\over g^{(E)}_{tt}}}=
\int_{r_s}^{(Q_1 Q_2 Q_3)^{1/3}}\!dr\, \sqrt{H_1 H_2 H_3} \approx 2\,\sqrt{Q_1 Q_2 Q_3\over r_s}
\label{traveltime}
\ee
We will show, in the next subsection, that this time matches the time computed from the dual CFT.

\subsubsection{The D1-D5-KK CFT}

The 3+1 dimensional system has a CFT description very similar to the 4+1 dimensional system. In the 4+1 dimensional case the D1 and D5 brane bound state was described by an `effective string' with total winding number $n_1 n_2$, and momentum excitations traveled on the components of this effective string. In the 3+1 dimensional case we can reach, by dualities, a description where we again have such an effective string. If we keep the three charges to be D1-D5-KK then we have an effective string with total winding number $n_1 n_2 n_3$. We can also consider M-theory compactified as
$M_{10,1}\rightarrow M_{3,1}\times T^6\times S^1$.  Let $z_1\dots z_6$ be the directions of $T^6$ and $y$ describe the $S^1$. Then we can wrap three sets of M5 branes along $(1234y), (1256y), (3456y)$, and the momentum excitations then run along the common direction $y$ \cite{klebanovtseytlin}. In each of these cases we get a CFT with central charge $6n_1n_2n_3$ where $n_1, n_2, n_3$ are the three charges.

Twist operators can link up the effective string into components of various lengths, just as in the D1-D5 case. For the {\it generic} state contributing to the entropy the component strings have winding number 
\be
k_i\sim \sqrt{n_1n_2n_3}
\ee
(This is the analogue of (\ref{vttone}).) The travel time for excitations around the component string is then
\be
\Delta t_{CFT}\sim  R\sqrt{n_1n_2n_3}
\ee
($R$ is the radius of the $S^1$.)

Let us compare  $\Delta t_{CFT}$  to $\Delta t_{sugra}$ (\ref{traveltime}). For D1-D5-KK we have
\be
Q_1={g\,\alpha'^3\,n_1\over 2 V {\tilde R}}\,,\quad 
Q_2={g\,\alpha'\,n_2\over 2 {\tilde R}}\,,\quad Q_3={{\tilde R}\,n_3\over 2}
\label{d1d5kkcharges}
\ee
From (\ref{rss}) we get
\be
r_s \sim{G_4\over R}
\ee
Substituting in (\ref{traveltime}) we get
\be
\Delta t_{sugra}\sim R\sqrt{n_1n_2n_3}
\ee
so that we find a relation analogous to (\ref{vtsix})
\be
\Delta t_{CFT}\sim \Delta t_{sugra}
\ee
We get a similar result from the system M5-M5-M5. (In this case the areas of the 2-tori $(56)$, $(34)$, $(12)$ are
$(2\pi)^2 T^{(1)}$, $(2\pi)^2 T^{(2)}$, $(2\pi)^2 T^{(3)}$
 and the radius $R$ of $S^1$ is $R=g\,\alpha'^{1/2}$.) We have
\be
Q_i={g\,\alpha'^{3/2}\,n_i\over 2 T^{(i)}}\,\,,\,\,i=1,2,3
\label{m5charges}
\ee
giving
\be
r_s\sim {G_4\over R}
\ee
and
\be
\Delta t_{sugra}\sim R\sqrt{n_1n_2n_3}\sim \Delta t_{CFT}
\ee

\subsection{The effect of compactification on horizons}

Here we discuss the subtlety mentioned above concerning compactifications of horizons.\footnote{We thank Rob Myers for pointing out this issue.}

Consider the naive geometry of D1-D5-P in 4+1 noncompact dimensions. The metric can be written in terms of harmonic functions
\bea
ds^2 &=& (H_1 H_5)^{-1/2}(-dt^2+dy^2+K(dt-dy)^2)\nonumber\\
&+& (H_1 H_5)^{1/2}(dz^2 + dr^2 + r^2 \,d\Omega_3^2)+\Bigl({H_1\over H_5}\Bigr)^{1/2} ds^2_X 
\label{lattice1}
\eea
\be
H_i = 1+{Q_i\over r^2 }\,\,,\,\,i=1,5; ~~~ K ={Q_p\over r^2 }
\label{lattice2}
\ee
\bea
Q_1 = {g\,\alpha'^3\,n_1\over V}\,,\quad
Q_5 = g\,\alpha'\,n_5 \,,\quad Q_p = {g^2\,\alpha'^4\,n_p\over R^2\,V}
\eea
with $(2\pi)^4V$ the volume of $X$ ($T^4$ or $K3$) and $R$ the radius of the $y$ circle.
If we compactify a transverse direction $z$ to a circle  of radius $\tilde R$, then the classical solution is given by
 an array of D1-D5-P black
holes placed along the $z$ direction with separation  $2\pi\,{\tilde R}$. This changes the harmonic functions to
\be
H_i = 1+\sum_{n=-\infty}^{\infty}{Q_i\over r^2 + (z- n 2\pi {\tilde R})^2}\,\,,\,\,i=1,5 ;~~~
K =\sum_{n=-\infty}^{\infty}{Q_p\over r^2 + (z- n2\pi  {\tilde R})^2}
\label{lattice3}
\ee
When $r\gg {\tilde R}$ the summations in (\ref{lattice3}) can be replaced by integrals 
\be
\sum_{n=-\infty}^{\infty}{1\over r^2 + (z- n 2\pi {\tilde R})^2}\,\,\rightarrow\,\,
{1\over 2\pi {\tilde R}}\int{dz'\over 
r^2 + (z-z')^2}= {1\over 2 {\tilde R}\,r}
\label{smearing}
\ee 
so that
\bea
H_i&\to& {\tilde H}_i={{\tilde Q}_i\over r}\,,\quad 
{\tilde Q}_i= {Q_i\over 2 {\tilde R}}\,\,,\,\,i=1,5\nonumber\\
K&\to& {\tilde K}={{\tilde Q}_p\over r}\,,\quad 
{\tilde Q}_p= {Q_p\over 2 {\tilde R}}
\label{smear}
\eea
and the solution (\ref{lattice2}) gives the naive metric (\ref{einstein3}) of three charges in 3+1 dimensions. However for
$r\lesssim {\tilde R}$ the  solution is  given by (\ref{lattice3}), and we see that it is not translationally invariant along  $z$.\footnote{A similar phenomenon was discussed in \cite{min} for the M5 brane geometry with a transverse direction compactified.}

By S,T dualities we can map the D1-D5-P charges to D1-D5-KK. Note that this is an exact duality  between charges, not just a classical map between solutions. Since all charges in this dualized solution  extend along the $z$ directions, we expect the geometry to be translationally {\it invariant} along $z$.

This difference in the translation properties of D1-D5-P and D1-D5-KK arises from taking  classical limits. For the true solution for D1-D5-P we should take the translationally noninvariant hole and let its center of mass be in  a wavefunction
which is constant in the $z$ direction. It is this complete solution that maps to D1-D5-KK under dualities. It is thus not possible to follow the horizon as a classical surface through the process of compactification and duality. But we do expect that in every dual description there should be a surface that measures the entropy of the 3-charge system, and this is what we have assumed in writing (\ref{rss}).   

Finally let us examine the location $r\sim \tilde R$ where the approximate solution (\ref{smear}) breaks down and the full solution (\ref{lattice3}) must be considered. If we compute the area of a boundary placed at $r=\tilde R$ then we find
\be
{A_{\tilde R}\over G_4}\sim \sqrt{n_1 n_5 n_p} 
\ee
We observe that the location where the approximate solution (\ref{smear}) breaks down is the same as the place where we
need to place a  horizon that would yield the entropy. Note that in the `fuzzball' picture  different microstates would end in different `caps' at this location.

\subsection{Summary}

We believe that the agreement $\Delta t_{CFT}\sim \Delta t_{sugra}$ for the three charge case is significant.  These travel times are not in any obvious way connected to the thermodynamic properties of the extremal hole. The similar agreement in the D1-D5 case arose from the relation between the finiteness of the CFT `component string' and the `capped' nature of the supergravity throat. Thus we are led to expect that for the three charge case there will be caps that reflect incident waves back after a time $\Delta t_{sugra}$. 

While we are not yet able to construct all microstates for three charges, some subfamilies of microstates  were constructed for D1-D5-P in 4+1 dimensions  in \cite{mss,gms1,gms2}. These geometries were found to be `capped'. Our computations above have suggested that D1-D5-P states with a transverse direction compactified end in `caps'; if we expand this compact direction back to infinite size we expect that we will continue to get a `capped' state rather than a smooth horizon.

\section{$R^2$ corrections to the geometries}

In a series of recent papers it has been argued that higher derivative terms in the low energy string effective action 
correct the 2-charge geometries -- the naive geometry with a zero area horizon is changed to one with a nonzero horizon satisfying $A/2G=S_{micro}$. We analyze such corrections in this section and argue that they should {\it not} alter the `fuzzball' picture of the microstates; i.e., we should not expect that such corrections replace the `capped' geometries by
a featureless smooth horizon.

In \cite{d} the correction terms in the action were derived from special 
geometry, based on earlier work \cite{dewitetal,osv}. In subsection 4.1 we look at  higher derivative terms in  string theory.  In 4.2 we observe that these terms give significant corrections to the {\it naive} geometry
 at the scale where the correction terms became significant in \cite{d}. Applying a similar analysis to the D1-D5 system on $K3\times S^1$   we  argue that  corrections would have to come from  string winding modes around  $S^1$.  In 4.3 we note that while this $S^1$ shrinks to zero in the {\it naive} geometry (and thus gives a diverging effect at $r\rightarrow 0$)
 there is no corresponding noncontractible $S^1$  in the {\it actual} capped geometries.

\subsection{Higher derivative corrections  in Type II string theory} 

In the subsections below we study $R^2$ terms in the dimensionally reduced theory. The reduction also generates scalars, and there are gauge fields present in the theory, but these are ignored in the discussion. We will use the order of magnitude of the $R^2$ terms to estimate the location where the correction terms become important. 

\subsubsection{$R^4$ terms in 10 dimensions}

The first corrections to the effective action of 10-dimensional type II theory involve terms with
eight derivatives: The $R^4$ terms \cite{gvz,gw,gs,gv,afmn,kp,gkkopp,kpa,pvw} 
together with terms involving derivatives of the dilaton, 
B-field and RR fields. We will restrict attention to the $R^4$ terms in what follows. The correction terms occur at string tree
level ($\alpha'$ effects) and  at string loop order ($g$ effects). 
The string tree level term is the same in type IIA and type IIB:
\be
S_{\rm tree}
={1\over 16\pi G_{10}} \int\!dx^{10}\,\sqrt{-g_{(S)}}\,e^{-2\phi}\Bigl[R_{(S)}+ 
{\alpha'^3\,\zeta(3)\over 3 \cdot 2^6}\,
(t_8 t_8 + {1\over 8}\epsilon_{10}\epsilon_{10})\,(R_{(S)})^4 \Bigr]
\label{action10Dtree}
\ee
while the 1-loop term distinguishes IIA and IIB\footnote{In the type IIB case there are also non-perturbative
corrections due to D-instantons; we ignore these since we will take $g$ small when studying D1-D5 in IIB.}:
\be
S^{(IIA)}_{\rm 1-loop}={1\over 16\pi G_{10}} \,{\alpha'^3\,\pi^2\over 9 \cdot 2^6} g^2 
\,\int\!dx^{10}\,\sqrt{-g_{(S)}}(t_8 t_8 - {1\over 8}\epsilon_{10}\epsilon_{10}) (R_{(S)})^4
\label{action10D1loopA}
\ee
\be
S^{(IIB)}_{\rm 1-loop}={1\over 16\pi G_{10}}\,{\alpha'^3\,\pi^2\over 9 \cdot 2^6} g^2 
\, \int\!dx^{10}\,\sqrt{-g_{(S)}}(t_8 t_8 + {1\over 8}\epsilon_{10}\epsilon_{10}) (R_{(S)})^4
\label{action10D1loopB}
\ee
($G_{10}=8\pi^6\,g^2\,\alpha'^4$.)
In the equations above we have used the following standard notation to denote index contractions:
\be
\epsilon_{10}\epsilon_{10} R^4\equiv \epsilon^{\alpha\beta\mu_1\nu_1\ldots\mu_4\nu_4}
\epsilon_{\alpha\beta\rho_1\sigma_1\ldots\rho_4\sigma_4}\,R^{\rho_1\sigma_1}_{\mu_1\nu_1}\ldots
R^{\rho_4\sigma_4}_{\mu_4\nu_4}
\ee
\be
t_8 t_8 R^4 \equiv  t_8^{\mu_1\nu_1\ldots\mu_4\nu_4}
t_{8\,\rho_1\sigma_1\ldots\rho_4\sigma_4}\,R^{\rho_1\sigma_1}_{\mu_1\nu_1}\ldots
R^{\rho_4\sigma_4}_{\mu_4\nu_4}
\ee
where the $t_8$ tensor is defined as ($F^{(i)}_{\mu\nu}$, $i=1,\ldots,4$ are any anti-symmetric tensors) 
\bea
&&t_8^{\mu_1\nu_1\ldots\mu_4\nu_4} F^{(1)}_{\mu_1\nu_1}\ldots F^{(4)}_{\mu_4\nu_4}\nonumber\\
&&= 8 (F^{(1)}_{\mu\nu}F^{(2)\,\nu\rho} F^{(3)}_{\rho\lambda} F^{(4)\,\lambda\mu}+
F^{(1)}_{\mu\nu}F^{(3)\,\nu\rho} F^{(2)}_{\rho\lambda} F^{(4)\,\lambda\mu}+
F^{(1)}_{\mu\nu}F^{(3)\,\nu\rho} F^{(4)}_{\rho\lambda} F^{(2)\,\lambda\mu})\nonumber\\
&& -2 (F^{(1)\,\mu\nu}F^{(2)}_{\mu\nu}F^{(3)\,\rho\lambda}F^{(4)}_{\rho\lambda}+
F^{(1)\,\mu\nu}F^{(3)}_{\mu\nu}F^{(2)\,\rho\lambda}F^{(4)}_{\rho\lambda}+
F^{(1)\,\mu\nu}F^{(4)}_{\mu\nu}F^{(2)\,\rho\lambda}F^{(3)}_{\rho\lambda})
\eea
Applied to the curvature tensor this contraction gives
\bea
t_8 t_8 R^4 &=& 12 (R_{\mu\nu\rho\sigma}R^{\mu\nu\rho\sigma})^2 + 
24 R_{\mu\nu\rho\sigma}R^{\rho\sigma\alpha\beta}R_{\alpha\beta\gamma\delta}R^{\gamma\delta\mu\nu}\nonumber\\
&-& 96 R_{\mu\nu\alpha\beta}{R^{\mu\nu}}_{\gamma\delta}R^{\rho\sigma\beta\gamma}{R_{\rho\sigma}}^{\delta\alpha}-
192 R_{\mu\nu\alpha\beta}R^{\mu\nu\alpha\gamma}R^{\rho\sigma\delta\beta}R_{\rho\sigma\delta\gamma}\nonumber\\
&+& 192 R_{\mu\nu\alpha\beta} R^{\rho\nu\gamma\beta} R_{\rho\sigma\gamma\delta}R^{\mu\sigma\alpha\delta}
+384 R_{\mu\nu\alpha\beta} R^{\rho\nu\gamma\beta}{R^{\sigma\mu}}_{\delta\gamma} {R_{\sigma\rho}}^{\delta\alpha}
\label{t8t8}
\eea

\subsubsection{Compactifications: Notation}

We will consider the reduction of these actions on $K3\times S^1$ and on $K3\times S^1\times {\tilde S}^1$. The volume of K3 is  $(2\pi)^4 V$ and $V_0$ is the value of $V$ at infinity. Similarly  $2\pi R$, 
$2\pi {\tilde R}$  are the lengths of $S^1$ and
${\tilde S}^1$ and $2\pi R_0$ and $2\pi {\tilde R}_0$ are the corresponding asymptotic values.  For the torus 
$T^2= S^1\times {\tilde S}^1$ we write
\be
T= R {\tilde R}\quad ~~\quad  \tau=i {{\tilde R}\over R}
\ee
to describe the volume  and complex structure modulus respectively. We will always let  the $T^2$ be rectangular and  the B-field on it be zero.

\subsubsection{Reduction to $M_{8,1}\times S^1$}

The dimensional reduction of the tree level terms is trivial (as mentioned above we will ignore the scalars arising from the reduction):
\be
S_{\rm tree}
={1\over 8 G_{10}} \int\!dx^{9}\,\sqrt{-g_{(S)}}\,R\,e^{-2\phi}
\Bigl[R_{(S)}+ {\alpha'^3\,\zeta(3)\over 3 \cdot 2^6}\,
(t_8 t_8 + {1\over 4}\epsilon_{9}\epsilon_{9})\,(R_{(S)})^4 \Bigr]
\label{action9Dtree}
\ee
The 1-loop terms, on the other hand,
receive additional contributions from string winding modes around $S^1$  \cite{gv} 
\bea
S^{(IIA)}_{\rm 1-loop}&=&{1\over 8 G_{10}}\,g^2\,{\alpha'^3\,\pi^2\over 9 \cdot 2^6}
\int\!dx^{9}\,\sqrt{-g_{(S)}}\,\Bigl[R
\,(t_8 t_8 - {1\over 4}\epsilon_{9}\epsilon_{9}) (R_{(S)})^4 \nonumber\\
&&\quad+{\alpha'\over R}\,
(t_8 t_8 + {1\over 4}\epsilon_{9}\epsilon_{9}) (R_{(S)})^4\Bigr]
\label{action9D1loopA}
\eea
\bea
S^{(IIB)}_{\rm 1-loop}&=&{1\over 8 G_{10}}\,g^2\,{\alpha'^3\,\pi^2\over 9 \cdot 2^6}
\int\!dx^{9}\,\sqrt{-g_{(S)}}\,\Bigl[R
\,(t_8 t_8 + {1\over 4}\epsilon_{8}\epsilon_{8}) (R_{(S)})^4 \nonumber\\
&&\quad + {\alpha'\over R}\,
(t_8 t_8 - {1\over 4}\epsilon_{9}\epsilon_{9}) (R_{(S)})^4\Bigr]
\label{action9D1loopB}
\eea
The terms  proportional to $R$ come from  naive
dimensional reduction of  10-D 1-loop terms and the  complete form including $\alpha'/R$ dependent terms
 is obtained by taking into account string winding modes. (The coefficients of the $\alpha'/R$ terms can be checked  by applying T-duality to the terms proportional to $R$.) We will apply the above actions to D1-D5 on $M_{4,1}\times K3\times S^1$ where the $S^1$ becomes small and so the $\alpha'/R$ terms are relevant.
 
\subsubsection{Reduction to $M_{7,1}\times S^1\times \tilde S^1$}

The tree level action is again just the dimensional reduction of the 10-D tree level term
\be
S_{\rm tree}
={\pi\over 4 G_{10}} \int\!dx^{8}\,\sqrt{-g_{(S)}}\,T\,e^{-2\phi}
\Bigl[R_{(S)}+ {\alpha'^3\,\zeta(3)\over 3 \cdot 2^6}\,
(t_8 t_8 + {1\over 4}\epsilon_{8}\epsilon_{8})\,(R_{(S)})^4 \Bigr]
\label{action8Dtree}
\ee
At  1-loop we can get contributions from string winding modes on $S^1, \tilde S^1$ as well as string worldsheet instantons wrapping  $T^2$. But in the cases that we wish to later work with (D0-D4 and D0-D4-F on $M_{3,1}\times K3\times T^2$) it is possible to keep
\be
{\tilde R}\gg R  , ~~~~~~
{\tilde R} R\gg\alpha'
\ee
everywhere. In this case we get contributions only from string winding modes on $S^1$, and the 1-loop contribution is just the further dimensional reduction of (\ref{action9D1loopA}),(\ref{action9D1loopB}):
\bea
S^{(IIA)}_{\rm 1-loop}&=&{\pi\over 4 G_{10}}\,g^2\,{\alpha'^3\,\pi^2\over 9 \cdot 2^6}
\int\!dx^{8}\,\sqrt{-g_{(S)}}\,\Bigl[T
\,(t_8 t_8 - {1\over 4}\epsilon_{8}\epsilon_{8}) (R_{(S)})^4 \nonumber\\
&&\quad+\alpha'\,\tau_2\,
(t_8 t_8 + {1\over 4}\epsilon_{8}\epsilon_{8}) (R_{(S)})^4\Bigr]
\label{action8D1loopA}
\eea
\bea
S^{(IIB)}_{\rm 1-loop}&=&{\pi\over 4 G_{10}}\,g^2\,{\alpha'^3\,\pi^2\over 9 \cdot 2^6}
\int\!dx^{8}\,\sqrt{-g_{(S)}}\,\Bigl[T
\,(t_8 t_8 + {1\over 4}\epsilon_{8}\epsilon_{8}) (R_{(S)})^4 \nonumber\\
&&\quad + \alpha'\,\tau_2\,
(t_8 t_8 - {1\over 4}\epsilon_{8}\epsilon_{8}) (R_{(S)})^4\Bigr]
\label{action8D1loopB}
\eea

\subsubsection{Reduction to $M_{3,1}\times K3\times T^2$}

The $K^3$ reduction of the $R^4$ terms in (\ref{action8Dtree}) and (\ref{action8D1loopA}-\ref{action8D1loopB}) will
produce both $R^2$ and $R^4$ terms in the effective theory on $M_{3,1}$. The correction terms in 
\cite{d, dkm} arise from $R^2$ terms \cite{hm}, which we now proceed to consider. These arise after two of the curvature factors are integrated over K3
\be
{1\over 32\pi^2}\int_{K3}\!dz^4\,\sqrt{g_{(K3)}}\, R^{(K3)}_{abcd}R_{(K3)}^{abcd} = 
c_2= 24
\label{k3}
\ee 
We get
\bea
&&\!\!\!\!\!\!\!\!\int_{M_{3,1}\times K3}\!dx^8\,\sqrt{-g_{(S)}}\,t_8 t_8 (R_{(S)})^4\\
&&\!\!\!\!\!\!\!\!\quad \to 24\,
\int_{K3}\!dz^4\,\sqrt{g_{(K3)}}\,R^{(K3)}_{abcd}R_{(K3)}^{abcd}\int_{M_{3,1}}
\!dx^4\,\sqrt{-g_{(S)}}\,R^{(S)}_{\mu\nu\rho\sigma}R_{(S)}^{\mu\nu\rho\sigma}\nonumber\\
&&\!\!\!\!\!\!\!\!\int_{M_{3,1}\times K3}\!dx^8\,\sqrt{-g_{(S)}}\,
{\epsilon_8 \epsilon_8\over 4} (R_{(S)}^4)\nonumber\\
&&\!\!\!\!\!\!\!\!\quad \to -24\,\int_{K3}\!dz^4\,\sqrt{g_{(K3)}}\,
R^{(K3)}_{abcd}R_{(K3)}^{abcd}\int_{M_{3,1}}
\!dx^4\,\sqrt{-g_{(S)}}\,\Bigl[R^{(S)}_{\mu\nu\rho\sigma}R_{(S)}^{\mu\nu\rho\sigma}-4 
R^{(S)}_{\mu\nu} R_{(S)}^{\mu\nu}+ R_{(S)}^2\Bigr]\nonumber
\eea
The terms containing $R, R_{\mu\nu}$  can be removed by a redefinition of 
variables \cite{mt}
\be
g_{\mu\nu}\to g_{\mu\nu}+ a\, R_{\mu\nu} + b \,g_{\mu\nu} R
\ee
and we find
\be
S^{(IIA)}_{R^2}={\pi\over 4 G_{10}}\,g^2\,\alpha'^3\,{8 \pi^4\over 3}\,c_2\,
\int\!dx^{4}\,\sqrt{-g_{(S)}}\,T\,R^{(S)}_{\mu\nu\rho\sigma}R_{(S)}^{\mu\nu\rho\sigma}
\label{action4Dr2A}
\ee
\be
S^{(IIB)}_{R^2}={\pi\over 4 G_{10}}\,g^2\,\alpha'^4\,{8 \pi^4\over 3}\,c_2\,
\int\!dx^{4}\,\sqrt{-g_{(S)}}\,\tau_2\,R^{(S)}_{\mu\nu\rho\sigma}R_{(S)}^{\mu\nu\rho\sigma}
\label{action4Dr2B}
\ee

The 4D Einstein metric $g_{(E)}$ is related to $g_{(S)}$ by
\be
g_{(E)}= g_{(S)}\,e^{-2\phi}\,{T\over T_0}\,{V\over V_0}
\ee
Thus the 4D Einstein actions of IIA and IIB including terms up to second order in the curvature are
\be
S^{(IIA)}_{4D}={4\pi^5 T_0 V_0\over G_{10}}\,\int\!dx^{4}\,\sqrt{-g_{(E)}}\Bigl[R_{(E)} + 
{c_2\over 6}\,{g^2\,\alpha'^3\over V_0}\,{T\over T_0}\,
R^{(E)}_{\mu\nu\rho\sigma}R_{(E)}^{\mu\nu\rho\sigma}\Bigl]
\label{action4DA}
\ee
\be
S^{(IIB)}_{4D}={4\pi^5 T_0 V_0\over G_{10}}\int\!dx^{4}\,\sqrt{-g_{(E)}}\,\Bigl[R_{(E)} + 
{c_2\over 6}\,{g^2\,\alpha'^3\over V_0}\,{\alpha'\over T_0}\,
\tau_2\,R^{(E)}_{\mu\nu\rho\sigma}R_{(E)}^{\mu\nu\rho\sigma}\Bigl]
\label{action4DB}
\ee

\subsubsection{Reduction to $M_{4,1}\times K3\times S^1$}

Finally we write down the reduction that we will need in studying the D1-D5 system. In 4+1 dimensions the Einstein metric is
\be
g_{(E)}=g_{(S)}\Bigl({R\over R_0} {V\over V_0}{\rm e}^{-2\phi}\Bigr)^{2/3}
\ee
Using the relations in subsubsection 4.1.3 we find that the reduced action in IIB theory is\footnote{This action is correct only  for small $g$ since we have not included the effects of windings of all other $(p,q)$ strings which enforce the full $SL(2,\mathbb{Z})$ symmetry of IIB string theory.}
\bea
S^{(IIB)}_{5D}&=&{2\pi^4 \over G_{10}}\int\!dx^{5}\,\sqrt{-g_{(S)}}\,R\,V\,e^{-2\phi}\Bigl[R_{(S)} + 
{c_2\over 6}\,{g^2\,\alpha'^4\over V\, R^2}\,e^{2\phi}\,
R^{(S)}_{\mu\nu\rho\sigma}R_{(S)}^{\mu\nu\rho\sigma}\Bigl]\\
&=&{2\pi^4 R_0 V_0\over G_{10}}\int\!dx^{5}\,\sqrt{-g_{(E)}}\,\Bigl[R_{(E)}
+ {c_2\over 6}\,{g^2\,\alpha'^4\over V_0\,R_0^2}\,\Bigl({V\over V_0} e^{-2\phi}\Bigr)^{-1/3}
\,\Bigl({R_0\over R}\Bigr)^{4/3}\,
R^{(E)}_{\mu\nu\rho\sigma}R_{(E)}^{\mu\nu\rho\sigma}\Bigl]\nonumber
\label{action5DB}
\eea

\subsection{Strength of the higher derivative corrections}

\subsubsection{The D0-D4 system}

Consider the D0-D4 system in type IIA theory compactified on $K3\times T^2$; this is the system studied in 
\cite{d}. At leading order in $\alpha'$ the 10-D metric and dilaton are (we parametrize 
$T^2=S^1\times {\tilde S}^1$ by $y$ and $z$)
\bea
&&\!\!\!\!\!\!\!\!\!ds^2=-(H_0 H_4)^{-1/2}dt^2 + (H_0 H_4)^{1/2} (dr^2+r^2 d\Omega_2^2)
+(H_0 H_4)^{1/2} (dy^2+dz^2)+
 \Bigl({H_0\over H_4}\Bigr)^{\!\!1/2}\!\!ds^2_{K3}\nonumber\\
&&\!\!\!\!\!\!\!\!\!e^{2\phi}=H_0^{3/2}\,H_4^{-1/2}\nonumber\\
&&\!\!\!\!\!\!\!\!\!H_0=1+{Q_0\over r}\,,\quad H_4=1+{Q_4\over r}
\eea
\be
Q_0 = {g\,\alpha'^{7/2}\,n_0\over 2 V_0 T_0}\,,\quad Q_4={g\,\alpha'^{3/2}\,n_4\over 2 T_0}
\label{chargesd0d4}
\ee
Since 
\be
e^{-2\phi}\, {T \,V\over T_0\,V_0} =H_0^{-3/2}\,H_4^{1/2} \, 
(H_0 H_4)^{1/2}\,{H_0\over H_4}=1
\ee
the 4-D Einstein metric $g_{(E)}$ is given by
\be
ds^2=-(H_0 H_4)^{-1/2} dt^2 + (H_0 H_4)^{1/2} (dr^2+r^2 d\Omega_2^2)
\label{naived0d4}
\ee
This metric is singular at $r=0$ and the area of the $S^2$ at $r=0$ vanishes.

It has been shown in \cite{d} that once the $R^2$ corrections to the type IIA action (\ref{action4DA}) are
taken into account the metric (\ref{naived0d4}) deforms into a metric with a non-singular horizon whose
associated entropy exactly reproduces the microscopic answer $4\pi\sqrt{n_0 n_4}$. We can get  a qualitative
understanding of this phenomenon by estimating the location $r_s$ at which the $R^2$ term in the action becomes
comparable with the Einstein-Hilbert term $R$. 
For $r\ll \sqrt{Q_0 Q_4}$ we find
\bea
&&R_{(E)}\sim (H_0 H_4)^{-1}\,\partial_r^2 (H_0 H_4)^{1/2}\sim {1\over r\,\sqrt{Q_0 Q_4}}\nonumber\\ 
&&R^{(E)}_{\mu\nu\rho\sigma}R_{(E)}^{\mu\nu\rho\sigma}\sim
\Bigl[(H_0 H_4)^{-1}\,\partial_r^2 (H_0 H_4)^{1/2}\Bigr]^2\sim {1\over r^2\,Q_0 Q_4} 
\label{vvtwo}
\eea
and 
\be
{T\over T_0}=(H_0 H_4)^{1/2}\approx {\sqrt{Q_0 Q_4}\over r}
\ee
We thus have that
\be
R_{(E)}\sim{g^2\,\alpha'^3\over V_0}\,{T\over T_0}\,R^{(E)}_{\mu\nu\rho\sigma}R_{(E)}^{\mu\nu\rho\sigma}
\ee
for 
\be
r \sim r_s={\alpha'^{3/2}\,g\over V_0^{1/2}}
\ee
The area of the spatial 2-sphere of radius $r_s$ in the metric $g_{(E)}$ is
(using $G_4={g^2\,\alpha'^4\over 8 V_0 T_0}$)
\be
A=4\pi\,\sqrt{Q_0 Q_4}~r_s \sim G_4\,\sqrt{n_0 n_4}
\label{vthir}
\ee
Thus a boundary placed at the location $r_s$ (where the higher derivative corrections become significant in the (naive) metric (\ref{vvtwo})) gives 
$A/G_4\sim S_{micro}$. This agreement  indicates that the correction terms we have written in string theory are the ones
relevant to the effect of \cite{d}.

It may be argued that we should compare the strengths of terms in the equations of motion and not in the action, but for the systems at hand these turn out to be equivalent. 
If we vary $g_{\mu\nu}$ in the 
 $R^2$ term in the type IIA 4D action (\ref{action4DA}) we get
\bea
&&\!\!\!\!\!\!\!\!\!\!\!\!\!\!\!\!\!\!\delta \int\!dx^{4}\sqrt{-g}\,T\,
R_{\alpha\beta\gamma\delta}R^{\alpha\beta\gamma\delta}=\int\!dx^{4}\sqrt{-g}\,
\delta g_{\mu\nu} \Bigl[{T\over 2} g^{\mu\nu} R_{\alpha\beta\gamma\delta}R^{\alpha\beta\gamma\delta}
-2 T  {R^\mu}_{\alpha\beta\gamma}R^{\nu\alpha\beta\gamma} \nonumber\\
&&\qquad\qquad\quad+ 2 T(\nabla^\mu\nabla^\nu R - 
2 \nabla_\alpha\nabla^\alpha R^{\mu\nu}+2 R^{\mu\alpha\beta\nu}R_{\alpha\beta} + 2 R^{\mu\alpha}{R^{\nu}}_\alpha)
\nonumber\\
&&\qquad\qquad\qquad\quad +8 \nabla_\alpha T (\nabla^\mu R^{\alpha\nu}- \nabla^\alpha R^{\mu\nu})+ 4 
(\nabla_\alpha\nabla_\beta T) R^{\mu\alpha\beta\nu}\Bigr]
\label{variationr2}
\eea
Note that for $r\ll \sqrt{Q_0 Q_4}$ all variables have the form $A/r^\alpha$, where $A,\alpha$ can depend on the choice of tensor indices on the component being examined. A little work then shows that for the metric (\ref{naived0d4}) we get 
(there is no sum over repeated indices)
\be
R_{\mu\nu\rho\sigma}\sim C_{\mu\nu\rho\sigma}(g_{\mu\rho}g_{\nu\sigma}-g_{\mu\sigma}g_{\nu\rho}) R\,,\quad 
\nabla_\mu \nabla_\nu T \sim C_{\mu\nu} g_{\mu\nu} R\,T
\label{vttthree}
\ee
where $C_{\mu\nu\rho\sigma},C_{\mu\nu}$ are constants of order unity. Using (\ref{vttthree})   in  (\ref{variationr2}) we find that the location where the $R^2$ terms become comparable to the $R$ terms in the equations of motion is the same location where the $R^2$ term  becomes comparable to the $R$ term in the action.

\subsubsection{The D1-D5 system}

We can do a similar analysis for the D1-D5 system in IIB theory on $K3\times S^1$. 
Consider the {\it naive} metric. At leading order in $\alpha'$  the  string metric and dilaton are
\bea
&&ds^2=(H_1 H_5)^{-1/2} (-dt^2+dy^2) + (H_1 H_5)^{1/2} (dr^2+r^2 d\Omega_3^2) +
\Bigl({H_1\over H_5}\Bigr)^{1/2}ds^2_{K3}\nonumber\\
&&e^{2\phi}={H_1\over H_5}\nonumber\\
&&H_1=1+{Q_1\over r^2}\,,\quad H_5=1+{Q_5\over r^2}
\label{vvone}
\eea
\be
Q_1 = {g\,\alpha'^3\,n_1\over V_0}\,,\quad Q_5=g\,\alpha'\,n_5
\label{chargesd1d5}
\ee
In this naive metric the radius of $S^1$ goes to zero at $r\rightarrow 0$
\be
R=R_0 (H_1 H_5)^{-1/4}\approx R_0 (Q_1 Q_5)^{-1/4}\,r
\ee 
The 5-D Einstein metric is
\be
ds^2_E=-(H_1H_5)^{-2/3}dt^2+(H_1H_5)^{1/3}(dr^2+r^2d\Omega_3^2)
\ee
 The curvature invariants of this metric are
\be
R_{(E)}\sim {(Q_1 Q_5)^{-1/3}\over r^{2/3}}\,,\quad R^{(E)}_{\mu\nu\rho\sigma}R_{(E)}^{\mu\nu\rho\sigma}\sim
{(Q_1 Q_5)^{-2/3}\over r^{4/3}}
\ee
and the place at which
\be
R_{(E)}\sim {g^2\,\alpha'^4\,\over V_0\,R_0^2}\,\Bigl({V\over V_0} e^{-2\phi}\Bigr)^{-1/3}\,
\Bigl({R_0\over R}\Bigr)^{4/3}\,R^{(E)}_{\mu\nu\rho\sigma}R_{(E)}^{\mu\nu\rho\sigma}
\ee
is
\be
r\sim r_s={\alpha'^2 g\over R_0 V_0^{1/2}} 
\ee
The area of the spatial 3-sphere of radius $r_s$ in the metric $g_{(E)}$
is (using $G_5={\pi\,g^2\,\alpha'^4\over 4 V_0 R_0}$)

\be
A=2\pi^2\sqrt{Q_1 Q_5}~r_s\sim G_5 \sqrt{n_1 n_5}
\ee
Thus a boundary placed at the location $r_s$ (where the higher derivative corrections become significant in the naive metric (\ref{vvone})) gives 
$A/G_5\sim S_{micro}$.
We see that  if we start with the naive geometry for D1-D5 on $K3\times S^1$ then we expect an effect similar to that found in \cite{d} for D0-D4 on $K3\times T^2$.

\subsection{Correction terms for `actual' metrics}

In section 2 it was shown that for the D1-D5 system on $T^4\times S^1$ the `actual' geometries dual to CFT microstates
were smoothly `capped'. For D1-D5 on $K3\times S^1$ we cannot dualize the FP solutions to get the D1-D5 solutions. Nevertheless
the solutions (\ref{six}) can also be interpreted as classical solutions to D1-D5 on $K3\times S^1$, since they do not involve the compact 4-manifold K3 or $T^4$. (The solutions where the F string vibrates in the $T^4$ direction will not map
to solutions for $K3\times S^1$.)

Let us now ask if the higher derivative corrections discussed above can change the `caps' of such solutions and generate a smooth horizon instead. One might immediately say that a smooth `capped' geometry  cannot change its topology by any corrections so as to generate a `horizon plus singularity'.  But we should note such a discontinuous change can happen if it is found that the correction terms {\it diverge} somewhere.

We have seen in subsections 4.1,4.2 that the correction terms for D1-D5 on $K3\times S^1$ arise from string winding modes on $S^1$, and that they diverge as this $S^1$ shrinks to zero. In the {\it naive} metric the $S^1$ indeed vanishes at $r\rightarrow 0$, and so the correction terms diverge at $r\rightarrow 0$.

In the {\it actual} geometries we see from (\ref{six}) that the coefficient of $dy^2$ vanishes where $\sqrt{1+K\over H}$ diverges, which happens along the curve $\vec x=\mu \vec F(v)$. At this curve we have seen that there is a coordinate singularity \cite{lmm}, and we have to change coordinates to understand the smoothness of the geometry. The metric near this curve is (\ref{vfourt}).
The change of coordinates
\bea
&&{\tilde r}^2=\rho, ~~
\tilde\theta={\theta\over 2}, ~~\tilde y={y\over 2{\tilde Q}}, ~~\tilde \phi=\phi-{y\over 2{\tilde Q}}
\nonumber\\ 
&&0\le\tilde\theta<{\pi\over 2}, ~~0\le \tilde y<{\pi R\over {\tilde Q}}=2\pi, ~~0\le\tilde\phi<2\pi
\eea
makes manifest the locally $R^4$ form of the metric \cite{grossperry}
\be
ds^2=4{\tilde Q}[d\tilde r^2+\tilde r^2(d\tilde\theta^2+\cos^2\tilde\theta d\tilde y^2+\sin^2\tilde\theta d\tilde\phi^2)]
\ee

Thus we see that the $y$ circle twists and mixes with the $S^3$, and this allows it to shrink to zero as the angular coordinate in a plane rather than as a noncontractible $S^1$. The twisting, in turn, 
is created by the off-diagonal terms $A_i, B_i$
in (\ref{six}) which are always nonzero in the actual microstates, and which are set to zero in the naive metric 
(\ref{naive}). 

We thus conclude that the higher derivative corrections remain bounded everywhere. So while they can correct the local details of the fuzzball state they cannot change the structure of the solution from a `fuzzball' to a metric with a smooth horizon enclosing a singularity. 

\section{Higher derivative corrections and the fuzzball picture of black holes}

Let us recall the `conventional' and `fuzzball' pictures for black holes:

\medskip

{\it Conventional picture:}\quad The horizon is a regular region of spacetime, and the interior of the hole
is `empty of information' except perhaps within a small neighborhood of the singularity.

\medskip

{\it Fuzzball picture:}\quad The information about the state of the hole is distributed throughout the interior of the `horizon'. Individual microstates do not have horizons; rather the `horizon' is the boundary of the region within which typical microstates
differ from each other. While the generic state is expected to have sizable quantum fluctuations, we can estimate the size of the typical 2-charge fuzzball by studying states which are described by classical geometries.  

\medskip

It has been argued in \cite{d} that higher derivative terms in string theory correct the metric of  D0-D4 on $K3\times T^2$,
generating a nonzero horizon. In this section we argue that while there may be several interesting (and yet to be understood) aspects of this computation, we should not conclude that they change the `fuzzball' states into a conventional black hole. 

\subsubsection{Off diagonal terms in the leading order solution}

We have seen that the 2-charge geometry at leading order has off-diagonal terms in the metric given by the functions
$A_i$ in (\ref{actualfp}), and the corresponding functions $A_i, B_i$ in (\ref{six}). If we wish to consider higher derivative corrections then we must start with an actual BPS solution which exists in the theory at leading order, and then add the corrections. In \cite{d} on the other hand we start with the naive metric (where the analogues of $A_i, B_i$  are set to zero), and then consider corrections. We cannot hope to get actual solutions of the theory by the latter procedure, since  at the place where we expect the horizon to come the terms $A_i, B_i$ are significant. In fact at the curve $\vec x=\mu\vec F(v)$ we have
\be
H^{-1} K\approx A_iA_i
\ee
(Both sides of this equation diverge at the curve, but the leading singularities are exactly equal.) The curve $\vec x=\mu\vec F(v)$ generically fills up a region that extends to the vicinity of the `horizon', and the off-diagonal terms are crucial in getting the geometries (\ref{six}) to `cap off'.

\subsubsection{Assumption of spherical symmetry}

In the computations of \cite{dewitetal, d, dkm, sen, hmr} we take the low energy effective action and find a 
solution assuming a {\it spherically symmetric} ansatz. The result is thus by construction spherically symmetric, and turns out to have a horizon. But if we did {\it not} impose the requirement of spherical symmetry, we have {\it other} solutions to the same low energy theory, with the {\it same} 
mass and conserved charges. Note that these latter solutions cannot become the spherically symmetric solution after higher derivative corrections are taken into account. To see this one can consider some nongeneric but easily understood solutions. For example, 
consider the special family (\ref{mm}) and let  $\gamma=1$. If we drop the factor $1$ in the harmonic functions we get
the near horizon geometry of the D1-D5 system. 
The change of coordinates \cite{balmm}
\be
\psi_{NS}=\psi-{a\over \sqrt{Q_1Q_5}}y, ~~~\phi_{NS}=\phi-{a\over \sqrt{Q_1Q_5}}t
\label{spectral}
\ee
brings this near horizon metric to the form 
\bea
ds^2&=&\sqrt{Q_1Q_5}[-{(r^2+a^2)\over {Q_1Q_5}}dt^2+{r^2\over {Q_1Q_5}}dy^2+{dr^2\over
r^2+a^2}\nonumber\\
&&~~~~~~~+(d\theta^2+\cos^2\theta d\psi_{NS}^2+\sin^2\theta
d\phi_{NS}^2)
+{1\over Q_5} dz_adz_a]
\label{innerns}
\eea
which is just ${(global}~AdS_3)\times S^3\times T^4$, a completely smooth geometry. This geometry cannot receive any significant corrections from higher derivative terms -- the curvature length scale everywhere is the $AdS$ radius or larger; moreover $AdS_3\times S^3$ is a symmetric space which cannot change its shape under corrections. (This solution has nonzero angular momentum, but that cannot affect the conclusion since we can  make similarly smooth solutions with no angular momentum -- we just take a vibration profile $\vec F(v)$ with wavelength comparable to the total length $L_T$ of the F string, but  which has no net rotation.)

If there are many solutions to the same supergravity equations with the same asymptotic charges, what is the significance of the spherically symmetric ansatz? In fact the solutions (\ref{six}) are smooth, with no horizon or singularity, so we must accept them as solutions to the leading order equations, and then correct them to higher orders. But the spherically symmetric solution leads 
to a horizon which must have a singularity inside; so we do {\it not} know if this is a solution in the full string theory.
From our understanding of 2-charge states we have seen that all microstates {\it must} break spherical symmetry (the F string cannot carry momentum P without a transverse deformation), so we do not expect that the spherically symmetric solution is a true solution of the 2-charge system.

\subsubsection{Orthogonality of the microstates}

We have constructed a continuous family of geometries for the FP system in the classical limit where the vibration of the F string was given by a classical profile. At the quantum level we must consider the individual energy eigenstates (\ref{veight}); coherent states made from these  energy eigenstates give the profiles $\vec F(v)$. The states (\ref{veight}) are orthogonal to each other, and their count gives the entropy. Now consider the gravity description of these states. In this description, after all corrections and quantum fluctuations are taken into account, the corresponding states must still be orthogonal. (We can imagine starting with an FP state at weak coupling, and follow the BPS state as we increase $g$ till we reach the gravity regime; orthogonal  states must remain orthogonal.) 

Our solutions (\ref{six}) all end in different caps, so it is easy to see that the semiclassical quantization of their moduli space will lead to orthogonal gravity states. But suppose higher derivative corrections changed all these solutions to the `conventional'  hole. Then the solutions are identical all the way down to the singularity, so we must have large differences between the microstates at the singularity to make the microstates  orthogonal. Thus it is incorrect to say that higher derivative corrections might eliminate the differences between microstates and give the `conventional' hole as an `averaged state'.  In any approach that argues for a conventional hole we must understand why the differences between solutions get moved from the entire `fuzzball' region to a small vicinity of the singularity.

\subsubsection{The geometric transition}

Setting aside the compact space $T^4$ or K3 (which plays no role in this discussion) we see that at spatial infinity we have a boundary $S^3\times S^1$. Here $S^3$ gives the angular directions and the D1,D5 branes extend along $S^1$. In the `matter' description, we ignore any backreaction on the geometry, and find that the $S^3$ shrinks to zero as $r \rightarrow 0$, while the $S^1$ remains nonzero. In the gravity description the $S^3$ stays finite, but the $S^1$ shrinks to zero \cite{lmm}. The off-diagonal terms appearing in the `actual' metrics are such that the $S^1$ mixes with the $S^3$, and thus there is no singularity where the $S^1$ vanishes. Such `geometric transitions' have been observed before in other examples of AdS/CFT duality \cite{gova,mn}.

We have seen that the higher derivative corrections do not diverge anywhere for the D1-D5 geometries. This happens because these corrections originate (for the D1-D5 system)   from winding modes of elementary strings. The `geometric transition' ensures that  the `actual' D1-D5 geometries do not have any nontrivial cycles which shrink to zero. (The $S^1$ (parametrized by $y$) did shrink to zero in the naive metric, but became the angular coordinate in a plane in the actual geometries.) Since these corrections do not become singular, they cannot change the `caps' into horizons and singularities.

\subsubsection{Summary}

We have argued that higher derivative corrections do not take the `fuzzball' geometries found at leading order and change them into the conventional black hole. It is interesting to see how the `actual' geometries avoid some of the properties of conventional holes. {\it If} we have a horizon, then analyzing the solution at the horizon gives a no-hair theorem. But in the actual geometries the horizon does not form, and so for the same mass and charges we have many solutions. Similarly, {\it if} we have a spherically symmetric solution then higher derivative corrections cause a nonzero horizon to appear. But the actual solutions all break spherical symmetry, in such a way that there is no noncontractible $S^1$, so the higher derivative corrections never diverge and horizons do not form. 

\section{Puzzles with higher derivative corrections}

It is remarkable that the higher derivative corrections for  D0-D4 on $K3\times T^2$ generate a horizon whose area gives exactly the Bekenstein entropy \cite{d}. Is this a general phenomenon that should extend to all similar cases? We find some puzzles in this regard, which we list below.

\medskip

(i) {\it D0-D4 on $T^4\times T^2$:} \quad We have obtained the $R^2$ terms in 4-D from $R^4$ terms in 10-D with two of the $R$ factors
being integrated on K3 as in (\ref{k3}). If we replace K3 by $T^4$ the corresponding integral vanishes, so we get no $R^2$ terms. How would we generate a horizon for D0-D4 on $T^4\times T^2$? We do get $R^4$ terms in 4-D, and perhaps these give a nontrivial effect, but this remains to be checked.

\medskip

(ii) {\it 3-charges in four dimensions:}\quad  We can take the compactification $K3\times T^2$ and consider {\it three} charges  instead of the two charges  in \cite{d}. For example we can take D0-D4-F, with the F wrapping a cycle of $T^2$. This system is T-dual to the D1-D5-P system of \cite{stromvafa}, with an additional transverse direction compactified. The leading order supergravity solution still gives a zero horizon area, and we can ask if the higher derivative corrections will generate a horizon whose area will give an entropy $2\pi\sqrt{n_1n_2n_3}$. In the appendix we estimate the corrections coming from $R^2$ terms in 4-D, and find that the effects they generate only give an entropy of order $\sim \sqrt{n_1n_2}$, which is much smaller than the expected value. It may be that there are other correction terms that need to be considered; this is an issue for further study.

\medskip

(iii) {\it The form of subleading terms in the entropy:}\quad
The computation of \cite{d} used the fact that with the compactification $K3\times T^2$ the entropy has a leading term quartic in the charges, and added to this is a subleading term quadratic in the charges  (the latter term gives the D0-D4 entropy). But it is not clear that for more general systems the entropy should have such a simple algebraic form for the subleading corrections. Consider the 3-charge D1-D5-P system on $K3\times S^1$. Because K3 has no odd cohomology, the entropy of the 3-charge states can be estimated by an index, the elliptic genus. This index counts the states of a CFT where the left movers are excited to a given level, but the right movers are in the ground state. But if we take the special case where the left movers are also at level zero, then we get the {\it two} charge states. The elliptic genus of K3 is defined as 
 \cite{dmvv} 
\be
\chi(K3;q,y)={\rm Tr}_{RR} \,(-1)^F\, y^{2\,J_L}\, q^{L_0-{c/24}}=\sum_{m\geq 0,l} c(m,l)\, q^{m}\, y^{l}
\ee
where the trace is over the RR Hilbert space of states, $F$ is the fermion number and $J_L$ is the generator
of the left-moving $U(1)$ current (normalized in such a way that $2 J_L$ is integer). Thus the coefficients 
$c(m,l)$ count the numbers of states in the RR sector of conformal dimension $m$ and $U(1)$ charge $l/2$.
The elliptic genus of  the D1-D5 CFT which has target space the symmetric product $S^N K3$ is given by the generating function 
\be
\sum_{N=0}^\infty p^N \,\chi(S^N K3; q,y)=\prod_{n>0,m\geq 0,l}{1\over (1-p^n q^m y^l)^{c(nm,l)}}
\ee
The coefficient of $q^0$ counts the 2-charge states, but it is unclear from this expression why the total entropy should
be a simple algebraic function containing  a 3-charge and a 2-charge term.

\section{Discussion}

We have computed travel times for 3-charge geometries in four dimensions, found agreement with the travel time in the dual CFT and observed that this agreement supports a picture of `caps' for three charge geometries. We have also argued that higher derivative correction for the D1-D5 system do not destroy the picture of `caps' since there are no nontrivial shrinking circles in the capped geometries and so the corrections  do not diverge anywhere.

What is the significance of the 2-charge geometry with horizon found in \cite{d}? We have seen that for very basic reasons no D1-D5 geometry is spherically symmetric, and by duality this fact extends to D0-D4. The solution of \cite{d} is on the other hand found by starting with a spherically symmetric ansatz. One might think that this symmetric solution denotes some kind of an average over all the microstates. There are two time scales in the black hole problem -- the short `light crossing time' across the hole and the much longer `Hawking evaporation time'. It is possible that over the long evaporation time the low energy radiation leaving the fuzzball can sense the details of the state and carry out its information, while for experiments over the shorter crossing time all microstates give similar results. If this is the case, then it may be that the conventional geometry of the hole can describe the latter short time experiments in some statistical way. To prove this however one would have to start with observables computed in actual microstates, examine the appropriate averages, and see what effective description emerges.

If we go to the Euclidean black hole then we have all the states of the hole moving in the closed time loop. The complete sum over all states is of course spherically symmetric, but this does not mean that we can replace the sum over the states in the loop by some `average state' running around the loop. Since the microstates differ from each other significantly in the interior of the fuzzball region $r< r_s$, it is not clear why a single classical geometry will be able to represent all of them as a `saddle point' for the complexified path integral.   

It will be of interest to see if the results of \cite{d} extend to 3-charges in four dimensions.   3-charge microstates constructed for special subclasses of states have turned out to be `capped' geometries \cite{mss,gms1,gms2}, and computations with supertubes \cite{supertubes} suggest that 3-charge states `swell up' to horizon sized `fuzzballs' just like 2-charge states. Understanding the relations between these different approaches should help in resolving black hole puzzles.

\section*{Acknowledgements}

We would like to thank  Atish Dabholkar, Rob Myers, Amanda Peet, Boris Pioline, Ashish Saxena, Ashoke Sen, Masaki
Shigemori, Yogesh Srivastava and Arkady Tseytlin for several helpful discussions. 
S.G. was supported by  an I.N.F.N. fellowship. The work of S.D.M  was
supported in part by DOE grant DE-FG02-91ER-40690.

\appendix
\section{$R^2$ corrections for 3-charge systems on $K3\times T^2$}
\renewcommand{\theequation}{A.\arabic{equation}}
\setcounter{equation}{0}

We consider the D0-D4-F system on $K3\times T^2$ (the D4 branes wrap K3, and the F strings wrap one cycle of $T^2$).
If we decompactify the other cycle of $T^2$, then this system is T-dual to the D1-D5-P system used in the entropy computation of Strominger and Vafa \cite{stromvafa}. Thus we expect the system to have an entropy
$2\pi\sqrt{n_0n_4n_1}$. In this Appendix we look at the $R^2$ terms in 4-D that we considered for the D0-D4 system. We find that for D0-D4-F the $R^2$ term in the action becomes comparable to the $R$ term at the same location as it did for the D0-D4 system, instead of the location where a three charge horizon would be expected. 

For the D0-D4-F system the 10D string metric and dilaton are
 \bea
ds^2&=&-(H_0 H_4)^{-1/2}H_1^{-1}\,dt^2 + (H_0 H_4)^{1/2}\,(dr^2 + r^2 d\Omega_2^2) \nonumber\\
&+& (H_0 H_4)^{1/2} H_1^{-1}\,dy^2+(H_0 H_4)^{1/2}\,dz^2+
\Bigl({H_0\over H_4}\Bigr)^{1/2}ds^2_{K3}\nonumber\\
&&e^{2\phi}=H_0^{3/2} H_4^{-1/2}H_1^{-1}\,,\quad H_i=1+{Q_i\over r}\,\,,\,\,i=0,4,1
\label{d0d4f1}
\eea
where the charges $Q_i$ are given by
\bea
&&Q_0 = {g\,\alpha'^{7/2}\,n_0\over 2 V_0\,T_0}\,,\quad Q_4 =
{g\,\alpha'^{3/2}\,n_4\over 2 T_0}\,,\quad
Q_1 = {g^2\,\alpha'^3 \,n_1 \over 2 V_0 {\tilde R}_0}
\eea

The 4D Einstein metric is 
\be
ds^2_{(E)}=-(H_0 H_4 H_1)^{-1/2}dt^2 + (H_0 H_4 H_1)^{1/2}(dr^2+r^2 d\Omega_2^2)
\label{d0d4f1einstein}
\ee
and its curvature invariants are of order
\be
R^{(E)}\sim {1\over \sqrt{Q_0 Q_4 Q_1\,r}}\,,\quad 
R^{(E)}_{\mu\nu\rho\sigma}R_{(E)}^{\mu\nu\rho\sigma}\sim {1\over Q_0 Q_4 Q_1\,r}
\label{curvatured0d4f1}
\ee

We wish to find the location $r\sim r_s$ where the $R$ and $R^2$ terms (\ref{action4DA}) are comparable. Using
(\ref{curvatured0d4f1}) and
\be
T=T_0\,(H_0 H_4)^{1/2}\,H_1^{-1/2}\approx T_0\,\sqrt{Q_0 Q_4\over Q_1}\,r^{-1/2}
\ee
we find
\be
r\sim r_s = {\alpha'^3 g^2\over Q_1 V_0}
\label{rsd0d4f1}
\ee
In the metric (\ref{d0d4f1einstein}) the area of the surface at $r=r_s$ is 
\be
A = 4\pi \sqrt{Q_0 Q_4 Q_1 \,r_s}\sim G_4 \sqrt{n_0 n_4}
\label{aread0d4f1}
\ee
This is of the same order as the area obtained in (\ref{vthir})  for the 2-charge  D0-D4 system, and is thus  much smaller than the area $8\pi\,G_4\,\sqrt{n_0 n_4 n_1}$ corresponding to the entropy 
of the 3-charge system. 

Let us also examine the corrections that arise from the fact that with higher derivative terms the entropy is not just given by the area of the horizon. 
The general formula for the entropy given in \cite{wald} is
\be
S=-{1\over 8 G_4}\,\int_{\Sigma} E^{\mu\nu\rho\sigma}\,\epsilon_{\mu\nu}\epsilon_{\rho\sigma}
\label{wald}
\ee
where $\int_\Sigma$ is the integral over the horizon computed with respect to the metric induced from $g_{(E)}$, 
$\epsilon_{\mu\nu}$ is the volume element in the space normal to $\Sigma$, normalized such that
\be
\epsilon_{\mu\nu}\epsilon^{\mu\nu}=-2
\ee
and 
\be
E^{\mu\nu\rho\sigma}={\delta L\over \delta R^{(E)}_{\mu\nu\rho\sigma}}
\ee
is the variation of the scalar Lagrangian $L$ with respect to the curvature, holding the metric fixed.
When $L$ is simply given by the Einstein-Hilbert term, (\ref{wald}) reduces to the usual area-entropy law
$S=A/4 G_4$. In our case the Lagrangian also has an $R^2$ term 
\be
L_{(R^2)}={c_2\over 6}\,{g^2\,\alpha'^3\over V_0}\,{T\over T_0}\,
R^{(E)}_{\mu\nu\rho\sigma}R_{(E)}^{\mu\nu\rho\sigma}
\ee
which leads to the following correction to the entropy
\be
S_{(R^2)}=-{1\over G_4}\,{c_2\over 24}\,{g^2\,\alpha'^3\over V_0}\,
\int_\Sigma {T\over T_0}\,R_{(E)}^{\mu\nu\rho\sigma}\epsilon_{\mu\nu}\epsilon_{\rho\sigma}
\ee
We use the curvature of the metric (\ref{d0d4f1einstein}) at the location $r=r_s$ to estimate the order of this correction.
We find
\be
S_{(R^2)}\sim {g^2\,\alpha'^3\over G_4 V_0}\,\sqrt{Q_0 Q_4\over Q_1\,r_s}\sim \sqrt{n_0 n_4}
\ee
which is the same order as the  entropy contributed from the area term at $r=r_s$. 

These computations suggest  that corrections to the entropy produced by the $R^2$ term are too small to account for the
entropy of the 3-charge hole. 
A similar analysis can be carried out for  D1-D5-KK on $K3\times T^2$.
As in the  D4-D0-F case, one finds that $R^2$ terms produce corrections to the entropy of order
$\sqrt{n_1 n_5}$, instead of expected 3-charge value $2\pi\sqrt{n_1 n_5 n_{KK}}$.


\begin{thebibliography}{99}

\bibitem{hawking}
S.~W.~Hawking,
Commun.\ Math.\ Phys.\  {\bf 43}, 199 (1975).


\bibitem{lm4}
O.~Lunin and S.~D.~Mathur,
Nucl.\ Phys.\ B {\bf 623}, 342 (2002), hep-th/0109154.


\bibitem{lm5}
O.~Lunin and S.~D.~Mathur,
Phys.\ Rev.\ Lett.\  {\bf 88}, 211303 (2002), hep-th/0202072.


\bibitem{mss}
S.~D.~Mathur, A.~Saxena and Y.~K.~Srivastava,
Nucl.\ Phys.\ B {\bf 680}, 415 (2004)
[arXiv:hep-th/0311092].

\bibitem{gms1}
S.~Giusto, S.~D.~Mathur and A.~Saxena,
Nucl.\ Phys.\ B {\bf 701}, 357 (2004)
[arXiv:hep-th/0405017].

\bibitem{gms2}
S.~Giusto, S.~D.~Mathur and A.~Saxena,
arXiv:hep-th/0406103.

\bibitem{lunin}
O.~Lunin,
JHEP {\bf 0404}, 054 (2004)
[arXiv:hep-th/0404006].


\bibitem{d}
A.~Dabholkar,
arXiv:hep-th/0409148.

\bibitem{dkm}
A.~Dabholkar, R.~Kallosh and A.~Maloney,
arXiv:hep-th/0410076.

\bibitem{sen}
A.~Sen,
arXiv:hep-th/0411255.

\bibitem{hmr}
V.~Hubeny, A.~Maloney and M.~Rangamani,
arXiv:hep-th/0411272.

\bibitem{stromvafa}
A.~Strominger and C.~Vafa,
Phys.\ Lett.\ B {\bf 379}, 99 (1996), hep-th/9601029.

\bibitem{senold}
A.~Sen,
Nucl.\ Phys.\ B {\bf 440}, 421 (1995)
[arXiv:hep-th/9411187];
A.~Sen,
Mod.\ Phys.\ Lett.\ A {\bf 10}, 2081 (1995)
[arXiv:hep-th/9504147].

\bibitem{susskind}
L.~Susskind,
arXiv:hep-th/9309145.

\bibitem{wald}
V.~Iyer and R.~M.~Wald,
Phys.\ Rev.\ D {\bf 50}, 846 (1994)
[arXiv:gr-qc/9403028];
V.~Iyer and R.~M.~Wald,
Phys.\ Rev.\ D {\bf 52}, 4430 (1995)
[arXiv:gr-qc/9503052].

\bibitem{lmm}
O.~Lunin, J.~Maldacena and L.~Maoz,
hep-th/0212210.

\bibitem{lm3}
O.~Lunin and S.~D.~Mathur,
Nucl.\ Phys.\ B {\bf 610}, 49 (2001), hep-th/0105136.

\bibitem{balmm}
V.~Balasubramanian, J.~de Boer, E.~Keski-Vakkuri and S.~F.~Ross,
Phys.\ Rev.\ D {\bf 64}, 064011 (2001), hep-th/0011217;
J.~M.~Maldacena and L.~Maoz,
JHEP {\bf 0212}, 055 (2002)
[arXiv:hep-th/0012025].

\bibitem{callanmalda}
C.~G.~Callan and J.~M.~Maldacena,
Nucl.\ Phys.\ B {\bf 472}, 591 (1996) hep-th/9602043.

\bibitem{jkm}
C.~V.~Johnson, R.~R.~Khuri and R.~C.~Myers,
Phys.\ Lett.\ B {\bf 378}, 78 (1996)
[arXiv:hep-th/9603061].

\bibitem{klebanovtseytlin}
I.~R.~Klebanov and A.~A.~Tseytlin,
Nucl.\ Phys.\ B {\bf 475}, 179 (1996)
[arXiv:hep-th/9604166].

\bibitem{min}
S.~Minwalla and N.~Seiberg,
JHEP {\bf 9906}, 007 (1999)
[arXiv:hep-th/9904142].
\bibitem{dewitetal}
G.~Lopes Cardoso, B.~de Wit and T.~Mohaupt,
Nucl.\ Phys.\ B {\bf 567}, 87 (2000)
[arXiv:hep-th/9906094];
T.~Mohaupt,
Fortsch.\ Phys.\  {\bf 49}, 3 (2001)
[arXiv:hep-th/0007195];
G.~Lopes Cardoso, B.~de Wit, J.~Kappeli and T.~Mohaupt,
JHEP {\bf 0012}, 019 (2000)
[arXiv:hep-th/0009234].

\bibitem{osv}
H.~Ooguri, A.~Strominger and C.~Vafa,
Phys.\ Rev.\ D {\bf 70}, 106007 (2004)
[arXiv:hep-th/0405146].

\bibitem{gvz}
M.~T.~Grisaru, A.~E.~M.~van de Ven and D.~Zanon,
Nucl.\ Phys.\ B {\bf 277}, 409 (1986).

\bibitem{gw}
D.~J.~Gross and E.~Witten,
Nucl.\ Phys.\ B {\bf 277}, 1 (1986).

\bibitem{gs}
D.~J.~Gross and J.~H.~Sloan,
Nucl.\ Phys.\ B {\bf 291}, 41 (1987).

\bibitem{gv}
M.~B.~Green and P.~Vanhove,
Phys.\ Lett.\ B {\bf 408}, 122 (1997)
[arXiv:hep-th/9704145].

\bibitem{afmn}
I.~Antoniadis, S.~Ferrara, R.~Minasian and K.~S.~Narain,
Nucl.\ Phys.\ B {\bf 507}, 571 (1997)
[arXiv:hep-th/9707013].

\bibitem{kp}
E.~Kiritsis and B.~Pioline,
Nucl.\ Phys.\ B {\bf 508}, 509 (1997)
[arXiv:hep-th/9707018].

\bibitem{gkkopp}
A.~Gregori, E.~Kiritsis, C.~Kounnas, N.~A.~Obers, P.~M.~Petropoulos and B.~Pioline,
Nucl.\ Phys.\ B {\bf 510}, 423 (1998)
[arXiv:hep-th/9708062].

\bibitem{kpa}
A.~Kehagias and H.~Partouche,
Phys.\ Lett.\ B {\bf 422}, 109 (1998)
[arXiv:hep-th/9710023].

\bibitem{pvw}
K.~Peeters, P.~Vanhove and A.~Westerberg,
Class.\ Quant.\ Grav.\  {\bf 18}, 843 (2001)
[arXiv:hep-th/0010167];
K.~Peeters, P.~Vanhove and A.~Westerberg,
Class.\ Quant.\ Grav.\  {\bf 19}, 2699 (2002)
[arXiv:hep-th/0112157].

\bibitem{hm}
J.~A.~Harvey and G.~W.~Moore,
Phys.\ Rev.\ D {\bf 57}, 2323 (1998)
[arXiv:hep-th/9610237].

\bibitem{mt}
R.~R.~Metsaev and A.~A.~Tseytlin,
Nucl.\ Phys.\ B {\bf 293}, 385 (1987).

\bibitem{grossperry}
D.~J.~Gross and M.~J.~Perry,
Nucl.\ Phys.\ B {\bf 226}, 29 (1983).

\bibitem{gova}
R.~Gopakumar and C.~Vafa,
Adv.\ Theor.\ Math.\ Phys.\  {\bf 3}, 1415 (1999)
[arXiv:hep-th/9811131].

\bibitem{mn}
J.~M.~Maldacena and C.~Nunez,
Phys.\ Rev.\ Lett.\  {\bf 86}, 588 (2001)
[arXiv:hep-th/0008001].

\bibitem{dmvv}
R.~Dijkgraaf, G.~W.~Moore, E.~Verlinde and H.~Verlinde,
Commun.\ Math.\ Phys.\  {\bf 185}, 197 (1997)
[arXiv:hep-th/9608096].

\bibitem{supertubes}
D.~Mateos and P.~K.~Townsend,
Phys.\ Rev.\ Lett.\  {\bf 87}, 011602 (2001)
[arXiv:hep-th/0103030];
R.~Emparan, D.~Mateos and P.~K.~Townsend,
JHEP {\bf 0107}, 011 (2001)
[arXiv:hep-th/0106012];
I.~Bena and P.~Kraus,
Phys.\ Rev.\ D {\bf 70}, 046003 (2004)
[arXiv:hep-th/0402144];
I.~Bena,
Phys.\ Rev.\ D {\bf 70}, 105018 (2004)
[arXiv:hep-th/0404073];
I.~Bena and N.~P.~Warner,
arXiv:hep-th/0408106;
B.~C.~Palmer and D.~Marolf,
JHEP {\bf 0406}, 028 (2004)
[arXiv:hep-th/0403025];
E.~G.~Gimon and P.~Horava,
arXiv:hep-th/0405019;
D.~Bak, Y.~Hyakutake and N.~Ohta,
Nucl.\ Phys.\ B {\bf 696}, 251 (2004)
[arXiv:hep-th/0404104];
D.~Bak, Y.~Hyakutake, S.~Kim and N.~Ohta,
arXiv:hep-th/0407253.

\end{thebibliography}
\end{document}